\documentclass[a4paper,11pt]{article}
\usepackage[utf8]{inputenc} \usepackage[T1]{fontenc}
\usepackage{times}
\usepackage[sf,pagestyles]{titlesec}
\titleformat{\section} {\normalfont\sffamily\bfseries} {\thesection}{1em}{}
\usepackage{amsmath}
\usepackage{amssymb}
\usepackage{graphicx}
\usepackage{cite}
\usepackage{xcolor}

\usepackage{enumitem}  % allows "resume"
\setlist[enumerate]{label={\hbox to 13mm{%
		\hbox to 0mm{\small\sf\textbf{Test}~\small\sf\textbf{\arabic{enumi}:}\hss}%
		   \hfill}},%
		leftmargin=15mm, labelwidth=13mm, labelsep=2mm, itemindent=0mm, itemsep=0.5mm}

\usepackage{psfrag}
\usepackage[bookmarksopen=true,bookmarksnumbered=true]{hyperref}
\usepackage[singlelinecheck=false,skip=-2mm,labelfont={bf,sf},font=small]{caption}

\long\def\comment#1{} 

\def\diffd{{\rm d}}  % the differential
\def\figureref#1{{{Figure~\ref{#1}}}} % references figures
\def\tableref#1{{{Table~\ref{#1}}}} % references table

\def\bullet{\leavevmode{\rule[0.14em]{0.24em}{0.24em}}\kern 0.33em}
\def\csrhd{\hbox to 0 pt{\hss\hbox to 13 pt{$\vartriangleright$\hss}}}

\usepackage[sf,pagestyles]{titlesec}
\titleformat{\section} {\normalfont\sffamily\bfseries} {\thesection}{1em}{}

\renewcommand{\thesection}{\arabic{section}}

\normalsize
\renewcommand{\baselinestretch}{1.1}

\usepackage{url}                % for url typesetting and linking
\renewcommand\url{\begingroup \urlstyle{same}\Url}
\usepackage{breakurl}

\makeatletter
\def\incepsfig{\@ifnextchar[{\@incepsf}{\@incepsf[tp]}}
\def\@incepsf[#1]#2#3#4{\@ifnextchar[{\@incepsfp[#1]{#2}{#3}{#4}}{\@incepsfp[#1]{#2}{#3}{#4}[]}}
\def\@incepsfp[#1]#2#3#4[#5]{%
\begin{figure*}[#1]%
  \captionsetup{width=150mm}
  \begin{center}%
	  \noindent\hbox to 140mm{\hss\scalebox{#3}{\hss\begin{psfrags}#5{\includegraphics{#2.eps}}\end{psfrags}\hss}\hss}%
  \end{center}%
  \caption{#4}\label{#2}
\end{figure*}}
\makeatother

\providecommand{\email}[1]{\hbox{#1}}
\providecommand{\institute}[1]{\footnote{#1}}

\let\oldbibliography\thebibliography
\renewcommand{\thebibliography}[1]{%
  \oldbibliography{#1}%
  \setlength{\itemsep}{-1pt}}

\makeatletter

\newcommand{\PreserveBackslash}[1]{\let\cstemp=\\#1\let\\=\cstemp} 
   \AtBeginDocument{%
   \ifx\arrayrulecolor\@undefined
      \def\arrayrulecolor#1#{\CT@arc{#1}}
      \def\CT@arc#1#2{%
         \ifdim\baselineskip=\z@\noalign\fi
         {\gdef\CT@arc@{\color#1{#2}}}}
      \let\CT@arc@\relax
   \fi}
   
\def\toprule{\arrayrulecolor{hks15}\noalign{\ifnum0=`}\fi
  \@aboverulesep=\abovetopsep
  \global\@belowrulesep=\aboverulesep %global cos for use in the next noalign
  \global\@thisruleclass=\@ne
  \@BTrule[\heavyrulewidth]}
\def\bottomrule{\arrayrulecolor{hks15}\noalign{\ifnum0=`}\fi
  \@aboverulesep=\aboverulesep
  \global\@belowrulesep=\belowbottomsep
  \global\@thisruleclass=\@ne
  \@BTrule[\heavyrulewidth]}
\def\midrule{\arrayrulecolor{hks15}\noalign{\ifnum0=`}\fi
  \@aboverulesep=\aboverulesep
  \global\@belowrulesep=\belowrulesep
  \global\@thisruleclass=\@ne
  \@BTrule[\lightrulewidth]}
\makeatother

\def\cstabhlinemuchdown{\rule[-3.5ex]{0mm}{4ex}} % lo spazio sopra e sotto \hline
\def\cstabhlinedown{\rule[-1.5ex]{0mm}{2ex}} % lo spazio sopra e sotto \hline
\def\cstabhlineup{\rule{0mm}{3.5ex}}

\newcommand{\blue}[1]{#1} % CS: makes nothing.
\usepackage{tikz}

\definecolor{lime}{HTML}{A6CE39}
\DeclareRobustCommand{\orcidicon}{%
	\begin{tikzpicture}
	\draw[lime, fill=lime] (0,0) 
	circle [radius=0.16] 
	node[white] {{\fontfamily{qag}\selectfont \tiny ID}};
	\draw[white, fill=white] (-0.0625,0.095) 
	circle [radius=0.007];
	\end{tikzpicture}
	\hspace{-2mm}
}

\foreach \x in {A, ..., Z}{%
	\expandafter\xdef\csname orcid\x\endcsname{\noexpand\href{https://orcid.org/\csname orcidauthor\x\endcsname}{\noexpand\orcidicon}}
}

\begin{document}        % Title page rewritten in Oct 2021

%\addcontentsline{toc}{section}{Testing a Conjecture On Quantum Chromodynamics} %

% Checked aloud.
\title{\Large\bfseries\sffamily Testing a Conjecture On Quantum Chromodynamics} %

\author{Christoph Schiller  \orcidA{} \institute{\ \ \ Motion Mountain Research, 81827 Munich, Germany. 
ORCID 0000-0002-8188-6282, \email{cs@motionmountain.net}}}

% NO \date command ADDS a date in US format.
% \date{} % This command puts the contents inside {} as the "date" - e.g. {} for nothing, or {Summer 2020} for Summer 2020.
\date{Revised November 2022}

\maketitle\normalfont\normalsize\sffamily  
 
\begin{abstract}  
%
% November 2022
%
\noindent\normalfont\normalsize% 
% ca 250 words.
% 
A Planck-scale model that includes quantum chromodynamics and goes
beyond it, is tested against observations.  The model is based on a
single fundamental principle.  Starting with Dirac's proposal describing 
spin $1/2$ particles as tethered objects, quarks and elementary fermions
are conjectured to be fluctuating rational tangles with unobservable
tethers.  Such tangles obey the free Dirac equation.  Classifying
rational tangles naturally yields the observed spectrum of elementary
fermions, including the six quark types and their quantum numbers.
\blue{Classifying tangle deformations naturally yields exactly three types
of gauge interactions, three types of elementary gauge
bosons, and the symmetry groups U(1), broken SU(2) and
SU(3).}  The possible rational tangles for quarks, leptons, Higgs and
gauge bosons allow only the observed Feynman diagrams.  The complete
Lagrangian of the standard model -- without any modification and
including the Lagrangian of quantum chromodynamics -- arises in a
natural manner.

Over 90 experimental consequences and tests about quark and gluon
behaviour are deduced from the single fundamental principle.  No 
consequence is in contrast with observations.  
The consequences of the strand conjecture include the complete quark model for
hadrons, the correct sign of hadron quadrupole moments, colour flux tubes,
confinement, Regge behaviour, running quark masses, correctly predicted hadron mass
sequences, the lack of CP violation for the strong interaction,
asymptotic freedom, \blue{and the appearance of glueballs.}
Two consequences differ from quantum chromodynamics.
First, the geometry of the strand process for the strong interaction
leads to an ab-initio estimate for the running strong coupling
constant.  Secondly, the tangle shapes lead to ab-initio lower and
upper limits for the mass values of the quarks.

\bigskip

\noindent\small Keywords: quantum field theory; quantum chromodynamics; 
strong coupling constant; quark masses; tangle model; strand conjecture. %

\bigskip

%\noindent\small PACS numbers: 12.38.-t (quantum chromodynamics), 12.10.-g (unified field theories and models).

\end{abstract}

\newpage\normalfont\normalsize

%-----------------------------------------------------------------------------
%-----------------------------------------------------------------------------
%-----------------------------------------------------------------------------

% November 2022
\renewcommand{\baselinestretch}{0.5}\normalsize\sf\normalfont\normalsize
\tableofcontents
\renewcommand{\baselinestretch}{1.1}\normalsize
\newpage\normalfont\normalsize
\newpage

%-----------------------------------------------------------------------------
%-----------------------------------------------------------------------------
%-----------------------------------------------------------------------------

\incepsfig{i-tm-QCDdiagram}{0.9}{The Feynman diagrams of the strong 
interaction allow expressing the main open questions of quantum
chromodynamics in the terms of perturbative quantum field theory: 
What happens at the interaction vertices?  What happens during
the propagation of quarks and gluons?}

%-----------------------------------------------------------------------------
%-----------------------------------------------------------------------------
\section{The open questions of quantum chromodynamics}   
\label{sec:QCDintro}

After the successful development of quantum chromodynamics in the twentieth century, the following questions remain:
\begin{quotation}
\noindent\csrhd What determines the number of different quarks? %

\noindent\csrhd What determines the gauge symmetry group SU(3), \blue{the 
lack of CP violation} and the mass gap of the strong interaction? %

\noindent\csrhd What determines confinement, the strong coupling constant and its running? %

\noindent\csrhd What determines the exact propagator of the quarks, including their mass values and their running? %

\end{quotation}
\noindent 
Answering these questions is necessary, for example, to understand 
fully the mass of the proton and the neutron, the masses of the nuclei, 
radioactive gamma-ray spectra, alpha decay, fusion, fission, the history 
of stars and of element formation. %
Simply put, one can ask: what happens at the fundamental interaction vertices 
of quantum chromodynamics shown in \figureref{i-tm-QCDdiagram}?
The present article proposes answers to these questions, \blue{as well as tests for checking them.}

% 24
Any candidate for a complete explanation of quantum chromodynamics has to realize a number of requirements. 
The explanation must derive the full elementary particle spectrum, the full gauge interaction spectrum, and 
also all the unexplained \emph{fundamental constants} of the standard model of particle physics: all coupling 
constants, all particle masses, all mixing angles and all CP violating phases. %
So far, it appears that the proposals found in the literature either failed to fulfil these requirements, 
or predicted new effects that failed to show up, or both, as explained by Shifman \cite{shifman}. %

% 32
The present article continues to explore and test a candidate for a unified description of quantum phenomena that 
has a simple foundation, agrees with experimental data, describes propagation, describes interaction vertices, 
predicts no physics beyond the standard model, and allows calculating the fundamental constants. %
The candidate, called the \emph{strand conjecture}, has been already explored and tested in particle physics in 
general, in quantum electrodynamics, and in general relativity, without finding any deviation from 
experiment \cite{cspepan,csorigin,csqed,csindian}. %
The strand conjecture is based on the idea by Dirac that particles are \emph{tethered} by fluctuating 
and unobservable strands -- assumed to be of Planck radius. %
The strand conjecture extends the idea by modelling particles themselves as tangles of strands.
The arising \emph{tangle model} for elementary particles naturally yields the particle spectrum, 
the gauge groups, and the Lagrangian of the standard model -- all without omissions, additions or modifications. %
For example, three generations of quarks, leptons and massive neutrinos arise naturally, including CKM and PMNS mixing. %
Overall, the tangle model predicts the lack of new physics.

% 28 
{In addition} to the full agreement with data, and \emph{in contrast} to the standard model, the strand conjecture 
yields results that go \textit{further}. %
The strand conjecture implies the possibility to calculate the fundamental constants of nature, in particular the gauge coupling 
constants, the elementary particle masses, as well as the various mixing angles and phases. %

% 30 - November 2022
To make the present article self-contained,  \blue{the first part summarizes} the strand conjecture and the tangle model 
for the Lagrangian and the particles of the standard model. %
These sections are taken from reference \cite{csqed}, which answered the open questions of quantum electrodynamics. %
The \blue{second, main part} deduces answers for all the open questions of quantum chromodynamics.
Throughout, a thorough list of numbered predictions and tests in the domain of the strong interaction are deduced. %
For fast reading, every section ends with a summary. \blue{The appendix summarizes how general relativity arises from strands.}

% 28
The present text argues that \emph{all} properties of the strong interactions -- including the exact QCD Lagrangian 
and its constants -- naturally follow from \emph{pictures of strands.} %
More exactly, QCD follows from the \emph{geometry of strands}.
In upcoming articles, the same will be argued for wave functions, the Schrödinger equation, entanglement, and for the weak interaction. 
This unconventional approach realizes the old dream that the description of nature at the fundamental 
level -- the Planck scale -- should be essentially \textit{algebraic}, not analytical, and that the description should, 
nevertheless, lead to continuous space and fields at all measurable scales. %

%-----------------------------------------------------------------------------
%-----------------------------------------------------------------------------
%-----------------------------------------------------------------------------
%-----------------------------------------------------------------------------
% NP
\phantomsection
\addtocontents{toc}{\vspace{1.5\baselineskip}}
\section*{Part I: From strands to the standard model}
\addcontentsline{toc}{section}{\textsf{\textbf{Part I: The strand conjecture and the tangle model}}}

%-----------------------------------------------------------------------------
%-----------------------------------------------------------------------------
\section{The origin of the strand conjecture} %
\label{sec:originsc}

For many years, Bohr presented quantum theory as a consequence of the smallest measurable action value $\hbar$ \cite{bohr}. %
Dirac included the energy speed limit $c$ into quantum theory. %
From around 1929 onward, Dirac showed the \emph{string trick} or \emph{belt trick} in his lectures. %
The belt trick, illustrated in \figureref{i-tm-belttrick}, describes the basic properties of spin $1/2$ -- the return to the original situation after a rotation by $4 \pi$, \blue{i.e., the double covering of SO(3) by SU(2)} -- as the result of \emph{tethering}. %
As Dirac explained \cite{gardner}, the belt trick also shows that a spin value below $\hbar/2$ is impossible.
Indeed, a smallest angular momentum value $\hbar/2$ also fixes the smallest observable action to the value $\hbar$.
Similarly, the fermion trick, illustrated in \figureref{i-fermiontrick}, describes the basic properties of fermions -- the return to the original situation ofter double particle exchange -- as the result of \emph{tethering}. %

%-----------------------------------------------------------------------------
%
\incepsfig{i-tm-belttrick}{1}{The \emph{belt trick} or \emph{string trick}: a double rotation, thus by $4\pi$, of a 
  tethered region, such as a belt buckle or a tangle core, is equivalent to no
  rotation -- if the tethers are allowed to fluctuate and untangle as shown.  %
  Untangling tethers is not possible after a particle rotation by only $2\pi$: %
       tangle cores with 4 or more tethers thus show the properties of {spin $1/2$} particles. %
  As a result, a tethered particle is able to rotate continuously. %
  In the strand conjecture, the trick also couples rotation and \emph{displacement} of the tangle core. % 
  The coupling of core rotation frequency and core displacement determines particle mass, as argued in Section \protect\ref{massest}. %
  (The figure is modified from reference \protect\cite{csorigin}.)} %
%-----------------------------------------------------------------------------

%-----------------------------------------------------------------------------
%
\incepsfig{i-fermiontrick}{1}{The \emph{fermion trick}: a double particle exchange of  
  tethered particles is equivalent to no
  exchange -- if the tethers are allowed to fluctuate and untangle as shown.
  Untangling tethers is not possible after single exchange. %
  Tangle cores with 4 or more tethers thus show the defining properties of {fermions}.} %
%-----------------------------------------------------------------------------

% 32 
In nature, all matter particles are observed to have spin $1/2$ and to be fermions. 
However, particle tethers are not observed.
In retrospect, Dirac's tether trick was the first hint that matter could be described with \emph{unobservable} extended constituents with \emph{observable} crossing switches. %
(A `crossing switch' is defined below.)
Fifty years later, in 1980, Battey-Pratt and Racey proved  \cite{bpr} that tethers not only imply spin $1/2$ behaviour and fermion behaviour, but that in addition, tethers imply the full free Dirac equation. %
In other terms, Battey-Pratt and Racey proved that Dirac's trick implies Dirac's equation.
In other terms, \emph{every quantum effect} can be seen as result of observable crossing switches of unobservable tethers. %

% 44 
Together with tethers, a further foundation of the strand conjecture is the realization that not only quantum theory and special relativity, but also general relativity is based on an invariant limit statement. For general relativity, the maximum force $c^4/4G$ \cite{sab,gib,cs2003,csmax,csprd1,csprd2},
or equivalently, the maximum power $c^5/4G$ or the maximum mass per length $c^2/4G$ can be used as defining principles: %
maximum force \blue{-- or any of the other limits --} implies Einstein's field equations.
Maximum force and has been thoroughly checked with thought experiments and real experiments. %
Maximum force is valid in nature -- and is independent of the validity of the strand conjecture.
\blue{The appendix provides more details on how strands comply with maximum force and how they reproduce gravitational mass, gravitons, and general relativity.}

% 44 
Taken together, the limit triplet -- special relativity's $c$, quantum theory's $\hbar$, and general relativity's $c^4/4G$ -- implies that there are no trans-Planckian effects in nature of any kind (as long as $G$ is substituted by $4G$ in all Planck quantities). %
These (corrected) Planck limits and all their consequences agree with all observations.
Therefore, the (corrected) Planck limits must also hold in a unified theory.
In fact, the simplest way to ensure the lack of trans-Planckian effects in a unified description of nature is to require that the smallest quantum of action $\hbar$, the smallest length $\sqrt{4G\hbar/c^3}$ and the smallest time $\sqrt{4G\hbar/c^5}$ hold everywhere and at every instant of time. %
The strand conjecture realizes these limits by design.

% 44
In summary, to ensure that the strand conjecture describes observations, all invariant limits -- such as $\hbar$, $c$, $c^4/4G$, $\sqrt{4G\hbar/c^3}$ and  $\sqrt{4G\hbar/c^5}$ -- and thus the lack of trans-Planckian effects, are taken as a basic property of nature. %
Together, unobservable tethers, observable crossing switches and invariant limits form the fundamental principle of the strand conjecture. %

%-----------------------------------------------------------------------------
%-----------------------------------------------------------------------------
\incepsfig{i-tm-fundamental-principle}{1}{%
The fundamental principle of the strand conjecture describes the simplest observation possible in nature, a fundamental event. %
In the strand conjecture, a fundamental event is Planck-sized, i.e., almost point-like. 
A fundamental event results from a \emph{skew strand crossing switch} -- the exchange of underpass and overpass -- at an approximate location in three-dimensional space and at an (approximate) instant in time. %
The strands themselves are unobservable, impenetrable, and best imagined as ropes or {cables} having Planck-size radius.  %
The crossing switch defines $\hbar$ as the unit of the physical action $W$. %
Both the Planck length and the Planck time arise, respectively, from the smallest and from the fastest possible crossing switch. %
The fastest crossing switch is discussed in references \protect\cite{csqed} and \protect\cite{csindian}.
%The crossing (switch) can also be taken as the strand realization of a qubit.% %
}[%
\psfrag{t1}{\small $t$}%
\psfrag{t2}{\small $t+\Delta t$}%
\psfrag{WW}{\small $W=\hbar$}%
\psfrag{DL}{\small $\Delta l \geq \sqrt{4\hbar G/c^3}$}%
\psfrag{DT}{\small $\Delta t \geq \sqrt{4\hbar G/c^5}$}%
\psfrag{KK}{\small $S=k$}%
]

%-----------------------------------------------------------------------------
%-----------------------------------------------------------------------------
\section{The strand conjecture and its fundamental principle} 
\label{sec:overviewsc}

The strand conjecture states: every system in nature -- matter, radiation, space, or horizon -- is made of strands that fluctuate at the Planck scale \cite{cspepan,csorigin}. %
\begin{quotation}
\noindent\csrhd A \emph{strand} is defined as smooth simple curved line -- a one-dimensional, 
open, continuous, everywhere infinitely differentiable subset of $ \mathbb{R}^3 $ or of a 
curved 3-dimensional Riemannian space, with trivial topology and without endpoints -- that is 
surrounded by a volume defined by a perpendicular disk of Planck radius $\sqrt{\hbar G/c^3}$ 
at each point of the line, whose shape is randomly fluctuating over time. 
The strand volume may not self-intersect; equivalently, the 
curvature radius of the smooth central line is never smaller than a Planck 
length. %
(The radius is observer-invariant.)
\end{quotation}
\noindent
In other terms, strands resemble thin flexible rubber ropes with a Planck-size radius 
-- with the additional property that they \emph{cannot be cut}, because they are \emph{not} made of parts. %
The thick lines in \figureref{i-tm-fundamental-principle} illustrate the flexible and locally cylindrical 
volume taken up by strands.
From the mathematical viewpoint, the above definition of a strand is similar to the definition of a 
rope that is used in knot theory for ropelength calculations \cite{old,canta}. %
Strands include \emph{shape fluctuations}.  Strands \emph{cannot} interpenetrate or intersect. %
From the physics viewpoint, strands  have \emph{no} observable properties.
But even though a strand is not observable, its topological tangling is.
This will become clear shortly. %

The tiny, but finite and impenetrable strand diameter visualizes the minimum length $\sqrt{4\hbar G/c^3}$ as the shortest distance possible between any two strand segments. %
By construction, strands imply that the minimum length is a lower limit for length observation and measurement. %

The strand conjecture makes the following claims:
\begin{quotation}
\noindent\csrhd Crossing switches -- the exchange of underpass and overpass -- \emph{determine the Planck units,} and in particular $\hbar$, as illustrated in \figureref{i-tm-fundamental-principle}. %

\noindent\csrhd Although strands are themselves {unobservable}, \emph{crossing switches are observable}, because of their relation to the quantum of action $\hbar$, the maximum speed $c$, the Boltzmann constant $k$ and the gravitational constant $G$. %

\noindent\csrhd Physical space is a \emph{network} of strands. 
Horizons are \emph{weaves} of strands. 
Particles are \emph{tangles} of strands. 

\noindent\csrhd Physical motion \emph{minimizes} the number of observable crossing switches of unobservable fluctuating strands. %

\end{quotation}
\noindent 
The first statement is called the \emph{fundamental principle}. %
The other statements follow from it; the last statement is the principle of least action.

Above all, the strand conjecture claims that \figureref{i-tm-fundamental-principle} contains all of physics: 
\emph{every modification} of the fundamental principle contradicts observation, \emph{every consequence} of the fundamental principle agrees with observation, and \emph{every observation} is described by the fundamental principle. %
In particular, the strand conjecture claims that \figureref{i-tm-fundamental-principle} contains quantum chromodynamics, as argued in the following. %

The term strand \emph{crossing} always implies a \emph{skew} crossing or \emph{apparent} crossing that only appears look like a cross in two-dimensional drawings. %
In three dimensions, every strand crossing always consists of a \emph{strand overpass} and a \emph{strand underpass}.
In three dimensions, strands are thus \emph{always} at a distance, as illustrated in \figureref{i-tm-fundamental-principle} and \figureref{i-tm-strand-crossing}. %
In particular, every crossing \emph{switch} -- the exchange of underpass and overpass -- always arises via strand deformations only. %
A full discussion of why crossing switches are observable -- and that nothing else is -- is given in reference \cite{csqed}: only crossing switches couple to the electromagnetic field. %

A crossing switch defines the simplest possible \emph{physical event.} 
In the strand conjecture, events are processes. 
The crossing switch is the most fundamental event and the most fundamental process.
All processes in nature are \emph{composed} of crossing switches; 
this includes macroscopic and microscopic \emph{motion} of matter and radiation, gauge \emph{interactions}, space \emph{curvature evolution}, and \emph{every measurement.}  %
The action of every moving macroscopic system consists of quanta $\hbar$, thus of crossing switches.

The origin of physical space from strands is discussed in reference \cite{csindian}. 
In a few words, the metric results from the density of crossing switches in the strand network that makes up space. %
Similarly, gravitational horizons \blue{and their entropy result from the crossing switches in strand weaves, as explained in the appendix.}
In this article, the strand network that makes up curved or flat space is not drawn in the figures.

In summary, every physical observable -- including length, mass or field intensities -- \emph{emerges} from combinations of crossing switches. %
Crossing switches define all physical units.
The crossing switch is the fundamental process in nature; every measurement, every motion and every interaction is a process built from crossing switches. %
The next sections show that continuous observables (and also wave functions) arise through strand fluctuations.
Then it is shown that particles and their properties, including quantum numbers, quark masses or coupling constants, arise through a combination of topological and geometric strand properties. %

%-----------------------------------------------------------------------------
\incepsfig{i-tm-strand-crossing}{1}{%
The geometric properties of a skew strand crossing -- a strand overpass and a strand underpass separated by a shortest distance  $s$ -- resemble those of wave functions. %
Both crossings and wave functions allow defining position, orientation, phase, and density.
For both cases, the \emph{absolute phase} value around the orientation axis can be chosen freely.
In contrast, for both cases, \emph{phase differences} due to rotations around the orientation axis are always uniquely defined.} %
%-----------------------------------------------------------------------------
\incepsfig{i-tm-fermion}{1}{%
In the strand conjecture, the wave function is due to \emph{fluctuating crossings}, and the probability density is due to \emph{fluctuating crossing switches} -- both after averaging. %
The phase of the wave function arises as the vector sum of all the crossing phases. 
Wave functions due to strand crossings form a Hilbert space. 
The tethers -- strands that continue up to large spatial distances -- lead to spin $1/2$ behaviour under rotations and to fermion behaviour under particle exchange. %
The tethers imply that tangle core rotation and core displacement are related; this allows defining a mass value.
The relation between core rotation and core displacement also implies that all fermion tangle cores move slower than light.} %
%-----------------------------------------------------------------------------

%-----------------------------------------------------------------------------
%-----------------------------------------------------------------------------
\section{Similar models} % 

Strands differ from superstrings. 
Superstrings have tension.  
Superstrings are located in ten- or higher-dimensional space. 
Also, superstrings are supersymmetric. 
Strands have none of these properties. 
Strands are neither bosonic nor fermionic; these properties apply 
only to tangles.
By definition, strands have no tension and live in three-dimensional space. 
As shown later on, strands explicitly exclude higher dimensions and supersymmetry.

Several authors have presented models that are also based on infinitely extended microscopic components.
Carlip \cite{carlip1,carlip2,carlip3} has described space as made of fluctuating lines in three spatial dimensions. 
Botta Cantcheff has published a similar model \cite{botta}.
Oriti has published an article based on a model that can be seen as strands glued together at their crossings \cite{oriti}.
Finally, the work of Asselmeyer-Maluga can be seen as a description of three-dimensional space \emph{between} the strands \cite{asselmeyer}, including its dynamics, instead of the strands themselves. %
So far, these authors have concentrated their research on gravity, and not yet fully included the gauge interactions, the elementary particles spectrum, or wave functions. %

%-----------------------------------------------------------------------------
%-----------------------------------------------------------------------------
\section[From strands to wave  functions and particles]{From strands to wave functions and particles\protect\footnote{Improved from reference \protect\cite{csqed}.}} % New in Jan 2022
\label{sec:overviewscwp}

This section summarizes the origin of wave functions. \blue{The central idea is:}
\begin{quotation}
\noindent\csrhd Strand crossings allow defining wave functions.
\end{quotation}
As illustrated in \figureref{i-tm-strand-crossing}, a skew strand crossing has the same mathematical properties that characterize a wave function: %
the geometry of a crossing allows defining density, position, orientation, and phase(s). %
(The \emph{density} is given by the inverse minimum distance; \emph{position} is the midpoint of the shortest distance segment $s$; \emph{orientation} and \emph{phase} are defined with suitable cross products and sums of the vector representing $s$ and of the two unit tangent vectors of the strands at the endpoints of $s$.) %
Geometrically, a crossing is described by \emph{one real number}, describing the minimum strand distance or density, and by \emph{four angles} defining the crossing geometry around the position of the crossing.  %
The geometric parameters of a crossing can thus be mapped to the \emph{two complex parameters} of the Pauli wave function -- or to (half of) those of a Dirac wave function. %
In particular, the \emph{phase of the wave function} of a particle arises as the sum of all crossing phases in the particle tangle, averaged over the fluctuations. %
The freedom in the definition of the absolute phase value at each crossing is at the origin of the freedom of gauge choice of the wave function \cite{cspepan,csorigin}. %

In the strand conjecture, all continuous fields arise through time averaging.
\begin{quotation}
\noindent\csrhd The averaging time is a few Planck times long.
\end{quotation}
The averaging of strand shape fluctuations is so fast that it has no other influence on observations.
Local wave function values, local field intensities and even space itself \cite{csindian} arise through time averaging of crossing and crossing switches. %

In the strand conjecture, all elementary fermions are \emph{rational} tangles, i.e., open tangles that are unknotted. % 
Only \emph{rational} tangles -- tangles that arise by moving tethers around in space, e.g. by twisting or braiding them -- allow reproducing the particle transformations that occur in interactions. %
No other topological structure is able to reproduce this: neither knots, nor links, nor prime tangles, nor virtual knots, nor loops, nor actual crossings, nor graphs. %
Equivalently, 
\begin{quotation}
\noindent\csrhd An elementary fermion is a rational tangle, made of unknotted but tangled tethers that fluctuate.
\end{quotation}
\noindent 
The rational tangles for the elementary fermions and bosons are illustrated in \figureref{i-tm-fermiontangles} and \figureref{i-tm-bosons} below.
For a tangle of fluctuating strands, the average crossing distribution is the \emph{wave function,} and the average crossing switch distribution is the \emph{probability distribution.} %
The connection is illustrated in \figureref{i-tm-fermion}. %
For a particle tangle, the average phase, the average density, and the two average spin orientation angles define the (first) two complex components of the Dirac wave function $\psi$ for a particle. %
For the \emph{mirror} tangle, the corresponding averages define the (last) two complex components of $\psi$, for the \emph{antiparticle}. %

%-----------------------------------------------------------------------------
\incepsfig{i-hise-belt}{0.92}{A simple way to visualize particle rotation for a lepton, which has six tethers, is the animation by Jason Hise \protect\cite{hisebelt}. %
  The rotating central cube symbolizes the tangle core.
  (Taken from reference \protect\cite{csorigin}.)}
%-----------------------------------------------------------------------------

As mentioned above, Dirac used a system equivalent to the one illustrated in \figureref{i-tm-belttrick} to demonstrate that a \emph{single} (tethered) tangle core behaves, under rotations, like a spin $1/2$ particle. %
\begin{quotation}
\noindent \csrhd When strands are imagined as ropes or cables that can be pulled at the ends (or at spatial infinity), the \emph{tangle core} is the region containing \emph{curved} strands.)
\end{quotation}

The fermion trick of \figureref{i-fermiontrick} confirms that a (tethered) tangle core behaves, under exchange, like a fermion.
Both results apply independently of the number of tethers, as long as their number is 3 or larger.
Videos that visualize both spin $1/2$ and fermion behaviour exist on the internet, produced by A. Martos \cite{fermionvideo}.
Still images from a further video, produced by J. Hise and illustrating particle spin as rotation, are shown in \figureref{i-hise-belt}.

In summary, in tethered tangles, averaging shape fluctuations lead to crossing densities, which reproduce wave functions.
Tangles also reproduce spin as tangle core rotation and particle exchange as tangle core exchange. %
The arising suspicion that \emph{every} quantum motion -- also translation, scattering and interactions -- can be described with tethered particles turns out to be correct. %

%-----------------------------------------------------------------------------
%-----------------------------------------------------------------------------
\section[Four ways from tethers to the free Dirac equation]{Four ways from tethers to the free Dirac equation\protect\footnote{Improved from reference \protect\cite{csqed}.}} % Rewritten completely in Feb 2020; March 2020
\label{sec:diror}

This section summarizes how, in 1980, Battey-Pratt and Racey proved \cite{bpr} that any \emph{tethered} massive quantum particle -- thus any small massive object with attached strands that leave up to spatial infinity -- is described by the Dirac equation for free particles. %
In other terms, Battey-Pratt and Racey assumed unobservable strands attached to a central mass and derived the Dirac equation. %

First of all, in the strand conjecture, the result of Battey-Pratt and Racey is extended. 
The massive particle itself is \emph{also} assumed to be made of strands: % 
     an elementary fermion is conjectured to be a (rational) tangle \emph{core} -- i.e., the tangled region of a tangle -- whose tethers reach up to large distances.  % Corrected Feb 2020
It will become clear later on why the tangle has to be \emph{rational}, i.e., formed only by braiding tethers, without any knots.

Battey-Pratt and Racey showed that a fermion moving (without interaction) through space can described by a constantly rotating mass (or tangle core) whose central position is advancing through space. %
The free motion of a tethered particle thus models Feynman's description of a quantum particle as an \emph{advancing} and \emph{rotating} arrow \cite{feynqed}: the arrow is the phase of the mass, i.e., of the tangle core. %
The relation between rotational and translational motion defines the \emph{inertial mass} of the particle.
In this way, the description of a particle with tethers also visualizes the description of the Dirac equation given by Hestenes \cite{hestenes,hestenes2,hestenes3}.
This description implies that advancing tangles with rotating cores are a model for fermion propagators.

Battey-Pratt and Racey started from the Dirac trick and showed that a tethered particle -- a tangle core in the present case -- defines the 4-component \emph{spinor} $\psi(x)$ in the following way (in the usual representation): %
\begin{quotation}
\noindent\csrhd Averaged over a few Planck times, the position of the \emph{center} of the core yields the \emph{maximum} of the probability density.

\noindent\csrhd At each position $x$, the upper two components of the spinor $\psi(x)$ are defined by the local average of finding, at that position, a tangle core with a specific orientation and phase.

\noindent\csrhd At each position $x$, the \emph{lower} two components of the spinor $\psi(x)$ are defined by the local average of finding, at that position, a \emph{mirror} tangle core with a specific orientation and phase.

\end{quotation}
\noindent 

Also in the strand conjecture, strands are not observable, but their crossing switches are.
The central connection -- between the belt trick and the Dirac equation -- discovered by Battey-Pratt and Racey is the following: 
\begin{quotation}

\noindent\blue{\csrhd Under relativistic boosts, the size and frequency of the belt trick 
behaves as expected from special relativity.}

\noindent\csrhd The belt trick implies the $\gamma^\mu$ matrices and their Clifford algebra, i.e., their geometric algebra properties \cite{hestenes,hestenes2,hestenes3}.

\noindent\csrhd The first two components of the  $\gamma^\mu$ matrices describe their effect on the tangle core, i.e., on the particle.

\noindent\csrhd The last two components of the  $\gamma^\mu$ matrices describe their effect on the mirror tangle core, i.e., on the antiparticle.
\end{quotation}
\noindent 	 
\blue{In simple terms, Battey-Pratt and Racey showed that tethered relativistic particles are described by the free Dirac equation.}
Their result can be rephrased in the following concise way:
\begin{quotation}
\noindent\csrhd The free Dirac equation is essentially a \emph{differential} version of Dirac's string trick, or belt trick. %
\end{quotation}
This is the basic discovery of Battey-Pratt and Racey. They wrote about it to Dirac, but he never answered.

\blue{In fact, the strand model extends the equivalence found by Battey-Pratt and Racey:} with the rational tangles given below for each elementary particle, particles and antiparticles can be transformed into each other by moving certain tethers with respect to the others. %
This is only possible with rational tangles, as illustrated below, in \figureref{i-tm-electronpositron}, \blue{and completes the equivalence of tethered structures with the Dirac equation.}

There are at least three other ways to see the connection between tethers and the Dirac equation. 
A second, equivalent way to understand the appearance of the free Dirac equation from strands is the following. 
The free Dirac equation 
\begin{equation}
	i \hbar \gamma^\mu \partial_\mu \psi = m \, c \, \psi 
\end{equation} 
arises from five basic properties: 
\begin{itemize}
	\itemsep0em 
\item[1.] The action limit given by $\hbar$, which yields wave functions $\psi$, 
\item[2.] The energy speed limit for massive particles given by $c$, which yields Lorentz transformations and invariance, 
\item[3.] The spin $1/2$ properties in Minkowski space-time, 
\item[4.] Particle--antiparticle symmetry, this and the previous point being described by the $\gamma^\mu$ matrices,
\item[5.] A particle mass value $m$ that connects phase rotation frequency and wavelength using the imaginary unit $i$. 
\end{itemize}
These five properties are necessary and sufficient to yield the free Dirac equation. 
(The connection between the $\gamma^\mu$ matrices and the geometry of spin was first made about a century ago by Fock and Iwanenko \cite{fock}.)
The tangle model of particles reproduces these five properties in the following way:
\begin{itemize}
	\itemsep0em 
\item[1.] All observables are due to crossing switches, which imply a minimum observable action $\hbar$ (see \figureref{i-tm-fundamental-principle}) and the existence of a wave function (see \figureref{i-tm-strand-crossing} and \figureref{i-tm-fermion}), %
\item[2.] Tangle cores are constrained to advance less than one Planck length per Planck time, thus less than $c$ (see \figureref{i-tm-belttrick}), %
\item[3.] Tangle core rotation connects rotation and displacement and generates a finite mass value $m$ much smaller than the Planck mass (see \figureref{i-tm-belttrick} and Section \ref{sec:parmass}), %
\item[4.] Tethering reproduces the spin $1/2$ properties for rotation, exchange and boosts, and thus introduces the $\gamma^\mu$ matrices (see \figureref{i-tm-belttrick}), with tangle and mirror tangle corresponding to particle and antiparticle, %
\item[5.] Tangle rotation through the belt trick corresponding to particle propagation.
\end{itemize}
As a result, the Dirac equation is a consequence of the tangle model.

A third way to see that the tangle model explains the existence of quantum motion uses the principle of least action.
When the least action principle is applied, the Lagrangian describes how to determine the value of the action.
Now, in the strand conjecture, \emph{action} is the number of crossing switches; each crossing switch produces an action value $\hbar$.
The principle of least action this simply becomes the \emph{principle of fewest crossing switches.}
For fermions, after suitable spatial averaging, this general idea leads to the free Dirac Lagrangian, and thus to the free Dirac equation. %

A fourth argument for the validity of the Dirac equation in the tangle model uses the derivation by Lerner \cite{lerner}.
He derives the Dirac equation from two basic properties: (1) from the conservation of spin current and (2) from Lorentz covariance.
Lerner showed that, together, these two properties imply the Dirac equation.
In the strand conjecture, first of all, the definition of spin using tethers implies, from simple topology, that the spin current is \emph{conserved}.
Second, the definition of spin using strands also implies the \emph{Lorentz covariance} of spin, i.e., the proper behaviour under rotations and boosts.
This second property was shown explicitly by Battey-Pratt and Racey \cite{bpr}.
In particular, Lorentz covariance arises because under boosts, strands change shape and their crossing density  -- the resulting wave function -- transforms as expected.
Both properties, spin current conservation and Lorentz covariance, are reproduced by the strand tangle model.
Therefore, fluctuating strands imply the free Dirac equation.

In summary, unobservable tethers allow deducing the free Dirac equation from the fundamental principle in at least four ways.
More precisely, the free Dirac equation results from the behaviour of crossings in fluctuating rational tangles. 
In other terms, strands visualize how the quantum of action $\hbar$ and the speed of light $c$ lead to the free Dirac equation.
The free Dirac equation in turn implies that the usual expressions for the fermion propagator follow from strands.

\blue{Written in terms of a Lagrangian for the case of quarks, the present section has thus shown that strands reproduce the free quark Lagrangian
\begin{equation}
    L_{\rm free\ quarks}= \sum_q^N \bar\psi_{qj}\,(i \hbar c \gamma^\mu \delta_{jk}\partial_\mu)\,\psi_{qk} - m_qc^2\bar\psi_{qj}\psi_{qj}
    \label{freelag}
\end{equation}
where $\psi_{qj}$ is the Dirac spinor of quark type $q$ with colour $j$ and mass $m_q$. In the following, it will become clear that strands also determine the mass values $m_q$ and the number  $N$ of quarks, as well as the terms for the gluon fields and for the strong interaction.}

\blue{A few remarks can help getting used to strands.} Both in nature and in the strand conjecture, the inability to observe action values below $\hbar$ leads to wave functions and probability densities. %
Both in nature and in the strand conjecture, the inability to observe speed values larger that $c$ leads to Lorentz invariance and the relativistic energy--momentum relation. %
Both in nature and in the strand conjecture, together with the mass and the spin $1/2$ properties due to tethers, the $\gamma^\mu$ matrices and the Dirac equation for a free particle arise, as explained by Simulik \cite{simulik, simulik2,simulik3}. %
Gauge fields will be included below.
Exactly like usual quantum theory, also the tangle model implies probabilities, Zitterbewegung, interference, a Hilbert space, contextuality, entanglement, mixing, decoherence all other quantum effects, as shown in detail elsewhere \cite{csvol6}. %

%-----------------------------------------------------------------------------
%-----------------------------------------------------------------------------
%-----------------------------------------------------------------------------
\section[Predictions about the free Dirac equation  and its limits]{Predictions about the free Dirac equation  and its limits\protect\footnote{Improved from reference \protect\cite{csqed}.}} % Rewritten in Feb 2020, March 2020, April 2020, Oct 2020
\label{sec:relqmpred}

Strands predict the lack of even the tiniest deviation from the free Dirac equation. %
Any such deviation would falsify the strand conjecture.
\begin{enumerate} % no resume - first one
   \item Observing a situation or an energy scale for which the free Dirac equation is not valid would falsify the strand conjecture.
\end{enumerate}
The prediction remains valid when gauge interactions are included, as deduced in the following sections.

In addition to the Dirac equation, the fundamental principle of the strand conjecture implies that every Planck unit (corrected by changing $G$ to $4G$) is an insurmountable \emph{local limit} to physical observables in the quantum domain. % \cite{csvol6}. % 
More precisely, the strand conjecture predicts the lack of any trans-Planckian effects. 
Therefore, 
\begin{enumerate}[resume]
   \item Observing an elementary particle whose energy is larger than $E_{\rm max}=\sqrt{\hbar c^5/4G} \approx 6.1 \cdot 10^{18}\,\rm GeV \approx 1 GJ$ would falsify the strand conjecture. The same applies to observations invalidating the (corrected) Planck limits for momentum, distance, action, acceleration, temperature, force, power, etc. %
\end{enumerate}
\noindent
\blue{In other terms, strands imply that the Dirac equation is only valid for length above the (corrected) Planck length an for energies below the (corrected) Planck energy.
In practice this is not a problem, because these limits cannot be approximated by many orders of magnitude.}

The tangle model implies that every elementary fermion, such as the electron, has an \emph{effective size} of the order of the Compton wavelength, explaining its wave properties and its spin properties.  %
Nevertheless, as shown in Section \ref{sec:fsttsi}, charged elementary fermions are \emph{effectively point particles} when probed by electromagnetic or other fields.  %
These two apparently contradictory requirements were spelled out clearly by Barut \cite{barut}. 
The tangle model realizes them, yielding  complete description of elementary particles. %
\begin{enumerate}[resume]
   \item Observing an elementary particle whose interaction are not local -- thus spread over a volume larger that $O(1)$ times the Planck volume -- would falsify the tangle model.
   \item Observing an elementary particle with a spin that is not an integer multiple of $\hbar/2 $, or an elementary particle that is neither a fermion or a boson, would falsify the tangle model.
\end{enumerate}

In summary, the description of the Dirac equation with strands is \emph{identical} to the usual one.
Strands imply that there are \emph{no} measurable deviations from the Dirac equation.
Nevertheless, rational tangles of strands imply \emph{a number of statements that go beyond the Dirac equation:} %
\begin{enumerate}[resume]
   \item The spectrum, the interactions, the quantum numbers, the masses and all the other elementary particle properties are \emph{not free parameters}, but are \emph{fixed} by their tangle structure.
\end{enumerate}
As a telling example, mass is predicted to be given by the frequency of the spontaneous belt trick.
The restrictions on the possible particles and interactions due to the tangle model are explored in the rest of this article.

%-----------------------------------------------------------------------------
\incepsfig{i-tm-bosons}{1}{The conjectured rational tangles for the elementary bosons of the standard model. %
Elementary boson tangles are made of one, two or three strands.
For each boson, the advancing tangle determines the spin value and the propagator.
The spin of bosons is 1, because any curved strand can rotate by $2\pi$ and return to its original shape.
All boson tangle cores rotate when propagating.
Photons and gluons are massless, and each are described by just a single tangle. 
The W and Z tangles are asymptotically planar.
The W, Z and Higgs are localizable and thus have mass; therefore they have additional, more complex tangles in addition to the simplest shown here (see text).
No further elementary boson (apart from the graviton \protect\cite{csindian}) is predicted to exist.
The W tangle is the only topologically chiral one, and thus the only electrically charged elementary boson.} %
%-----------------------------------------------------------------------------

%-----------------------------------------------------------------------------
%-----------------------------------------------------------------------------
%-----------------------------------------------------------------------------
\section[Predictions about the spectrum of elementary particles]{Predictions about the spectrum of elementary particles\protect\footnote{Improved from reference \protect\cite{csqed}.}} % Rewritten in Feb 2020, March 2020, April 2020, Oct 2020
\label{sec:partpred}

This section summarizes how strands lead to the observed spectrum of elementary bosons and fermions.
The details were already presented elsewhere \cite{cspepan,csorigin}.

%-----------------------------------------------------------------------------
\incepsfig{i-tm-fermiontangles}{0.80}{The simplest conjectured tangles for each elementary fermion.
All elementary fermion tangles are made of two or three strands.
Elementary fermions are rational, i.e., unknotted tangles.
The cores are localizable and realize the belt trick. %
Tangles generate spin $1/2$ behaviour, positive mass values, and exactly three generations. %
The fermion tangle structure leads to Higgs coupling, as illustrated in \protect\figureref{i-lagr2}.
At large distances from the tangle core, the four tethers of the quarks follow the axes of a tetrahedron. %
At large distances from the core, the six tethers of the leptons follow the three coordinate axes. %
Neutrino cores are simpler when seen in three dimensions: they are twisted triples of strands. %
Neutrino cores are chiral but not topologically chiral: thus they are electrically neutral.
The tangles of the electron, the muon and the tau are topologically chiral, and thus electrically charged.
All massive particles have additional, more complex tangles in addition to the one shown here (see text). %
No additional elementary fermions appear.}
%-----------------------------------------------------------------------------

\emph{Elementary bosons} can consist of one, two or three strands. %
More strands imply \emph{composite} systems. %
The Reidemeister moves from mathematical knot theory yield the known gauge groups -- as explained below -- and thus suggest that one-stranded bosons correspond to photons, two-stranded bosons to the $W_1$, $W_2$ or $W_3$, and three-stranded bosons to gluons. %
After symmetry breaking, when two-stranded boson tangles incorporate a vacuum strand, they yield the three-stranded W and Z bosons. %
The complete overview of boson tangles is given in \figureref{i-tm-bosons}. %
Photon and gluon tangles are massless, because they can move unhindered by tethers, whereas the W and the Z boson have mass. 
No additional elementary gauge boson appears possible: higher number of strands are not possible in elementary particle tangles, and more complex tangling of the strands of bosons would yield fermion behaviour instead. %
\begin{enumerate}[resume]
\item {The discovery of additional elementary gauge bosons -- including those due to a higher gauge symmetry, due to a fifth force, or due to supersymmetry --  would falsify the tangle model.} %
\end{enumerate}
\noindent
The Higgs boson is a \emph{braid}, made of three strands. %
For all massive particles, Higgs braids can be added to the tangle core, as illustrated in \figureref{i-3-generations} and \figureref{i-lagr2}. 
Every massive particle -- fermion or boson -- is thus described by an \emph{infinite family} of tangles that contain a simple core, that core plus one Higgs braid, that core plus two braids, etc. %
The mass value is influenced by this -- single or multiple -- Higgs boson addition. %
\figureref{i-lagr2} shows that the Higgs couples to itself; it is thus massive.
Because no addition of a Higgs braid to cores of massless elementary particles is topologically possible, 
     massless elementary particles are described by a \emph{single} tangle. %
\begin{enumerate}[resume]
\item {The discovery of additional Higgs bosons would falsify the tangle model.} %
\end{enumerate}
\noindent
\emph{Elementary fermions} can consist of two or three strands. %
One-stranded particle tangles cannot have spin $1/2$ nor have mass because the belt trick is not applicable to them. %
Two-stranded fermions are quarks, three-stranded fermions are leptons. %
The \emph{simplest} specific tangles for each fermion are given in \figureref{i-tm-fermiontangles}. %
All fermions have \emph{additional} tangles: each fermion is described by an infinite \emph{family} of tangles that contains the simplest tangle core, that simplest core plus one Higgs braid, the simplest core plus two braids, etc. %
Strands thus reproduce Yukawa coupling.

Both quarks and leptons are limited to \emph{three generations} by the coupling to the Higgs and by the three-dimensionality of space. %
\figureref{i-3-generations} shows how the infinite class of quark-like braids is split into six infinite families, corresponding to the three generations. %
The infinitely many possible quark tangles consist of $6+6$ separate infinite tangle and mirror tangle families in which each tangle and each mirror tangle differs from the other by one or several Higgs braids. %
Each infinite tangle series corresponds to a quark or to an antiquark.

The quark--tangle assignments in \figureref{i-tm-fermiontangles} reproduce the quark model of 
hadrons \cite{csorigin,csvol6}, including quark chirality and the correct retrodiction of which mesons violate CP symmetry; 
quark tangles also reproduce all meson and hadron mass sequences. %
This will be shown in detail below.

Together, the lepton tangle assignments and the quark tangle assignments reproduce the weak interaction and its violation of parity.
Particle mixing is explained in reference \cite{cspepan}.
The tangle assignments for neutrinos explain their handedness and their small mass.
Additional elementary fermions are not possible: such tangles cannot have one strand, nor can the have four or more strands; and if a candidate fermion tangle is made of two or three strands, it is already included in the infinite families. %
\begin{enumerate}[resume]
      \item {The discovery of any additional elementary fermion, i.e., of any fermion beyond those of standard model -- including a fourth generation or a superparticle -- would falsify the tangle model.} %
\end{enumerate}
\blue{Together, \figureref{i-tm-fermiontangles} and \figureref{i-3-generations}
fix the number $N$ of quarks as
\begin{equation} 
       N=6 \;\;.
\end{equation}
The quark number $N$ appears in the QCD Lagrangian given in expression (\ref{freelag}); its explanation was one of the unsolved problems of the standard model.}

Because elementary particles are made at most of 3 strands, we also have
\begin{enumerate}[resume]
      \item There is no elementary particle with spin 3/2. 
	  The discovery of such a particle would falsify the tangle model. 
\end{enumerate}

In summary, in the strand conjecture, tangle classification leads to the fermion and boson spectrum observed in nature. %
As explained in the next section, every observed quantum number is due to a \emph{topological} property of particle tangles, more precisely, to a \emph{topological invariant.} %
(In contrast, the fundamental constants, such as mass or coupling strength, are due to averaged \emph{geometric} properties of tangles.)
The appearance of the gauge groups from the boson tangles is summarized below.
\begin{enumerate}[resume]
\item The discovery of any new elementary particle of any kind -- such as anyons, axions, supersymmetric partners, or any new elementary dark matter particle -- would falsify the tangle model. %
\end{enumerate}

%-----------------------------------------------------------------------------
\incepsfig{i-3-generations}{0.86}{Quarks consist of six two-stranded tangle families, each with an infinite number of tangles.
The six families define the three generations. %
The same happens with anti-quarks, which are represented by mirror tangles. %
In the strand conjecture, the number of quark generations is thus due to the tangle structure and its Yukawa coupling to the Higgs.
About the d quark, see more details later on in the text.} %
%-----------------------------------------------------------------------------

%-----------------------------------------------------------------------------
\incepsfig{i-tm-electronpositron}{1}{The simplest tangle of the electron (left) can be continuously deformed into the simplest tangle of the positron (right), provided that tethers `ends' are allowed to change position.} %
%-----------------------------------------------------------------------------

%-----------------------------------------------------------------------------
\incepsfig{i-gauge}{0.9}{The three Reidemeister moves classify all the possible deformations of tangle cores \protect\cite{reidem}. 
The moves also determine the generators of the observed gauge interactions, determine the generator algebra, and thus fix the three gauge groups \protect\cite{cspepan,csvol6}. %
Every group generator \emph{rotates} the region enclosed by a dotted circle by the angle $\pi$. %
The full gauge group arises by generalizing these local rotations to arbitrary angles, as
explained in Section~\protect\ref{sec:gintpred}. (Improved from reference \protect\cite{csorigin}.)}
%-----------------------------------------------------------------------------

%-----------------------------------------------------------------------------
%-----------------------------------------------------------------------------
%-----------------------------------------------------------------------------
\section{Predictions about the structure of the elementary fermions, including quarks} 
\label{sec:parttruc}

In the strand conjecture, elementary fermions are \emph{rational tangles} \cite{cspepan}.
Every elementary fermion is described by an infinite tangle family defined by (1) a simplest rational tangle, plus (2) all those rational tangles that arise when Higgs braids of three strands are successively inserted at one end. %
This is visualized for quarks in \figureref{i-3-generations}.
The rational tangle model of fermions can be tested by exploring its consequences.

The simplest tangles for fermion families can be classified.
The possible simplest tangles for each fermion are given in \figureref{i-tm-fermiontangles}. %
\begin{enumerate}[resume]
   \item Discovering any additional elementary fermion or elementary fermion generation would invalidate the tangle model.
\end{enumerate}
\noindent
Every quantum number is a \emph{topological} property of the corresponding particle tangle. 
Quantum numbers naturally are integers or simple fractions.
The tangle of each elementary particle determines helicity and parity P from its static behaviour and rotation behaviour after reflection.
The tangle determines spin from the rotation behaviour of the tethered tangle core.
The tangle determines baryon and lepton number from the number of tethers.
The tangle determines flavour quantum numbers, from the quark content, i.e., from the core topology. %
Electric charge is due to tangle core chirality \cite{csorigin,csqed}. 
Electrically neutral tangles have negative charge parity C if their mirror image is rotated by $\pi$.
Colour charge is explored below.
Weak charge was mentioned in references \cite{cspepan} and \cite{csorigin} and will be presented in detail in a future article on the weak interaction.
\begin{enumerate}[resume]
   \item The discovery of forbidden values of quantum numbers, or of the non-conservation of baryon number or lepton number (in processes described by perturbative quantum field theory), would falsify the tangle model. %
\end{enumerate}
Also non-perturbative effects that contradict the tangle model would invalidate it. This issue is still subject of research.
\begin{enumerate}[resume]
   \item The discovery of new quantum numbers, such as supersymmetry's R-parity, would falsify the tangle model. 
\end{enumerate}
Fermion mass is determined by the frequency of the belt trick. 
In the case of quarks, the details are explored below.

The operation of \emph{charge conjugation} C switches all crossings in a tangle.
The operation has the expected properties: it has only two eigenvalues, the \emph{C parities} $+1$ and $-1$; it transforms a particle into its antiparticle (defined by its mirror tangle); eigenstates must be neutral particles; all other quantum numbers change sign under charge conjugation; finally, C parity is multiplicative: the value of composite particles is the product of the values of the components. Mass remains unchanged.

The operation of \emph{parity inversion} P takes the mirror image of a tangle.
The operation has the expected properties: it has only two eigenvalues, the even and odd \emph{P parities} $+1$ and $-1$; it changes helicity; it is multiplicative: the value of composite particles is the product of the values of the components.

\figureref{i-tm-electronpositron} shows that for the spinning electron, the simplest tangle core is essentially a continuously rotating triangle formed by its three strands.
The three crossings each yield a third of the elementary charge.
Depending on the spatial approximation, the three rotating crossings can be seen as forming a rotating torus, or, at larger distance, as forming a rotating vortex. %
At even larger distance, the electron is a point particle.
In this sense, the tangle structure resembles several other electron models proposed in the past, such as those by Hestenes \cite{hestenes2,hestenes3}. %
\begin{enumerate}[resume]
   \item Discovering a non-vanishing elementary fermion size -- in interactions -- would invalidate the strand conjecture.  %
   \item Discovering any contradiction between tangle properties and observed particle properties would invalidate the strand conjecture.  %
   For example, discovering a new energy scale in high energy physics or any other deviation from the high energy desert would invalidate the strand conjecture. %
\end{enumerate}

Fermion tangles allow a smooth transition (by moving tethers against each other) between tangle and mirror tangle. %
For the electron and the positron, the transition is illustrated in \figureref{i-tm-electronpositron}. %
The tangle model is thus also able to visualize that for a given Dirac spinor $\psi(x)$, the ratio between electron and positron probability density can \emph{vary} from one position $x$ to another. %
It may well be that the rational tangle model of the electron is the \emph{only model} that allows this visualization.
(For example, non-rational tangles do not show mixing of particles and antiparticles.
Non-rational tangles also cannot reproduce particle pair production or flavour change.)

%-----------------------------------------------------------------------------
 \incepsfig{i-quarkrotation}{0.91}{The continuous spinning rotation of a down quark, with its four tethers, visualized with an animation
 by Jason Hise. The full animation can be watched at \protect\url{www.motionmountain.net/research.html\#dqt}.}
%-----------------------------------------------------------------------------

Quark tangles, iluustrated in \figureref{i-tm-fermiontangles}, show the same general behaviour as electron tangles.
In the tangle model, spin is tangle core rotation. 
The belt trick allows such rotations without end.
\figureref{i-quarkrotation} shows the rotation of the spinning down quark. 
Quark tangles have spin $1/2$.

Quark tangles and antiquark tangles are related by tether exchanges.
The tangle model reproduces the electric charge of quarks.
Only six generations of quarks are possible.
Quark tangles (except for the simplest d quark tangle), like the electron tangle, are \emph{chiral}.
As shown in detail below, the quark tangles reproduce quark mass, colour, confinement and the quark model of hadrons.

In summary, the tangle model proposes a \emph{specific structure} of tangles that explains the properties and quantum numbers of elementary particles.
\blue{No unobserved particle is predicted to exist.
In fact, the strand prediction is even stronger:}
\begin{enumerate}[resume]
   \item Discovering any elementary particle constituents that describe observations but \emph{differ} from tangles -- such as preons, rishons, ribbons, Möbius bands, strings, graphs, mebranes, tetrahedra, prequarks, knots, tori, qubits etc. -- would \textit{falsify} the strand conjecture. %
\end{enumerate}
\noindent 
It has to be noted that \emph{it may be} that some fermion tangles are wrongly 
assigned in \figureref{i-tm-fermiontangles}, while the strand conjecture as a whole remains correct. %
In particular, the lepton tangles need critical scrutiny.

%-----------------------------------------------------------------------------
%-----------------------------------------------------------------------------
%-----------------------------------------------------------------------------
\section{Predictions about gauge interactions} 
\label{sec:gintpred}

This section summarizes earlier results \cite{cspepan,csorigin} showing that gauge interactions are related to \emph{tangle core deformations.}
The general connection is illustrated in \figureref{i-gauge}:
deformations of a tangle core \emph{modify} the phase of the corresponding particle. % 
\blue{The phase of a quantum particle can thus change in two ways. First, the phase can change during free propagation, when the tangle core rotates as a whole while advancing. Secondly, the phase can change when the tangle core is deformed by the absorption of a gauge boson, which changes is shape and thus its phase.}
For example, tangle deformations reproduce how an externally applied magnetic field modifies the phase of an electron wave function through the absorption of virtual photons. %

The deformations of three-dimensional objects form gauge groups. %
In 1926, Reidemeister showed that every tangle core deformation is composed of three basic types: \emph{twists, pokes} and \emph{slides} \cite{reidem}. %
Today, the three deformations are also called the first, second and third \emph{Reidemeister moves}. %
The moves have a property that is not widely known \cite{cs2003,csmax,cspepan}:
\begin{quotation}
   \noindent \csrhd {Tangle core deformations -- given by Reidemeister moves -- determine the observed gauge groups U(1), (broken) SU(2) and SU(3).}
\end{quotation}
In particular, the gauge group U(1) arises because twists, the first Reidemeister move that forms a loop by deforming a short segment of a strand, can be generalized to arbitrary angles \blue{$\beta$.
Twists can also be concatenated: a twist by an angle $\beta_1$ followed by one by an angle $\beta_2$ is the same as a twist by an angle $\beta_1 + \beta_2$. %
Above all, a double twist, i.e., a twist by an angle $2 \pi$, can be rearranged to yield no twist at all.
A twist by an angle $\beta$ can thus represented by the complex number
\begin{equation}
   {\rm e}^{i \beta} \;\;,
\end{equation}
so that the non-trivial topology of the Lie group U(1) arises \cite{csqed}.} %
In short,
\begin{quotation}
   \noindent \csrhd Twists, or first Reidemeister moves, define and reproduce the gauge group U(1).
\end{quotation}
Electric charge is defined in Section \ref{sec:6elch} as $1/3$ of the (signed) sum of chiral crossings.  
\emph{Electric fields} are volume densities of virtual photons, i.e., of twists. \emph{Magnetic fields} are flow densities of twists. 
As an automatic consequence \blue{of the definition of electric charge,} only massive tangles can be electrically charged \cite{csqed}.
\begin{enumerate}[resume]
   \item Discovering a massless but electrically charged elementary particle would falsify the strand conjecture. %
   \item Observing the slightest deviation from U(1) or QED would falsify the strand conjecture. %
\end{enumerate}
This short summary of quantum electrodynamics is explored and tested in detail in reference \cite{csqed}.  
\blue{It is found that twists lead to Maxwell's equations and reproduce 
all observed consequences of electromagnetism, including the anomalous g-factor.}

The gauge group SU(2) arises because pokes, the second Reidemeister move, can be seen as localized rotations by the angle $\pi$ around the three coordinate axes.
\blue{\figureref{i-gauge} illustrates that the circular region defined by two strands can rotate in three different directions. 
Each rotation direction leads back to itself after two full turns.
This belt-trick behaviour of rotations by the angle $\pi$ in three perpendicular directions is known to lead to}
an SU(2) Lie algebra \cite{cspepan}. %
\blue{Indeed, the three pokes $\tau_{x}$, $\tau_{y}$, and $\tau_{z}$ are local rotations of two strand segments by~$\pi$; strands illustrate that their squares yield~$-1$ (four pokes in a row are like no poke) and that there is a cyclic anticommutation relation between them.  
Simple experimenting with strands yield the poke multiplication table
\begin{equation}
 \let\l\tau 
\begin{array}{c|ccc} 
  \cdot & \l_x    &    \l_y &    \l_z  \\  
\hline % instead of midrule 
\l_x    &   -1    &  i\l_z  & -i\l_y   \\ 
\l_y    & -i\l_z  &  -1     &  i\l_x   \\ 
\l_z    &  i\l_y  & -i\l_x  &   -1     \\ 
\end{array}
\label{weaktab}
\end{equation}
Therefore, pokes generate an SU(2) algebra.} 
The generalization of these pokes -- or local rotations by $\pi$ -- to arbitrary angles yields the full SU(2) Lie group. %
\begin{quotation}
   \noindent \csrhd Pokes, or second Reidemeister moves, define and reproduce the gauge group SU(2).
\end{quotation}
Strands further imply that only massive fermions can exchange weak bosons.
This is observed.
Due to the tangle structure of fermions, \emph{maximal parity violation} arises: %
    parity violation occurs because the core \textit{rotations} due to spin $1/2$ interfere with the core \textit{deformations} due to the group SU(2) of the pokes of the weak interactions \cite{cspepan,csvol6}. %
In addition, \emph{SU(2) breaking} arises: %
    a vacuum strand is included in the massless bosons, leading to the W and Z boson tangles \cite{cspepan,csorigin}. %
\begin{enumerate}[resume]
   \item Discovering any deviation from broken SU(2) or from the known (electro-) weak interaction properties of the standard model would falsify the strand conjecture.
\end{enumerate}
As explained below, also quark mixing and neutrino mixing arise from the strand tangle model.

Finally, previous articles \cite{cspepan,csorigin} have shown: %
\begin{quotation}
   \noindent \csrhd Slides, or third Reidemeister moves, define SU(3).
\end{quotation}
\blue{Also the SU(3) Lie algebra arises because slides, or third Reidemeister moves, can be seen as local rotations by $\pi$, and because the concatenation of these local rotations reproduce the multiplication of generators of SU(3).
This is explained below, in Section~\ref{sec:basicgluons}, using  \figureref{i-gluon} and Table~1.
The full gauge group SU(3) arises when these rotations by $\pi$ are generalized to arbitrary angles. %
Color charge is given by the orientation of the three-ended side of a quark tangle in space.}
Color fields are densities of virtual gluons. 
As a result, usual quantum chromodynamics arises.
\begin{enumerate}[resume]
   \item Observing the slightest deviations from SU(3), from the Lagrangian of quantum chromodynamics, or from the known strong interaction properties would falsify the strand conjecture.
\end{enumerate}

In summary, because there are only three Reidemeister moves, strands predict only three gauge interactions. 
The classification of strand deformations produces 
the observed gauge groups. 
In particular, strands predict the lack of any other gauge interaction:
\begin{enumerate}[resume]
   \item Discovering a larger gauge group, or a different gauge group, or a new elementary gauge boson, or a different interaction beyond the standard model (or any effect beyond general relativity) would falsify the strand conjecture. %
\end{enumerate}
So far, all experiments confirm this restrictive prediction \cite{pdgnew}.

%-----------------------------------------------------------------------------
\incepsfig{i-lagr1}{0.8}{The interaction vertices allowed by fermion and boson topologies 
   imply the complete Lagrangian of the standard model (part one). (Improved from reference \protect\cite{csorigin}.)}
%-----------------------------------------------------------------------------

%-----------------------------------------------------------------------------
\incepsfig{i-lagr2}{0.8}{The interaction vertices allowed by fermion and boson topologies  
   imply the complete Lagrangian of the standard model (part two). 
   (Improved from reference \protect\cite{csorigin}.)}
%-----------------------------------------------------------------------------

%-----------------------------------------------------------------------------
%-----------------------------------------------------------------------------
%-----------------------------------------------------------------------------
\section[Predictions about the standard model]{Predictions about the standard model\protect\footnote{Improved from reference \protect\cite{csqed}.}} % New in Feb, March 2020
\label{sec:predsm}

As just explained, the strand conjecture leads to restrictions on the possible elementary fermions and on the possible elementary bosons and interactions.
If the tangle topologies are explored, all Feynman diagrams of the standard model (with massive Dirac neutrinos and PMNS mixing) are recovered.
This result from references \cite{cspepan} and \cite{csorigin} is summarized in \figureref{i-lagr1} and \figureref{i-lagr2}.
The topology of tangles does not allow additional Feynman vertices.

In more detail, the argument is the following.

\begin{itemize}
\itemsep 0em 
\item The fundamental principle implies that particle can be modelled as tangles.
\item Free fermion tangles obey the free Dirac equation. They obey the corresponding propagator (see Section \ref{sec:diror}) and thus are described by the Dirac Lagrangian. %
\item Rational tangles determine the \emph{fermion} spectrum -- exactly three generations of quarks (made of two strands) and of leptons (made of three strands) -- with the observed particle properties: spin, charges, representations, other quantum numbers, and masses. The elementary fermion spectrum  is illustrated in \figureref{i-tm-fermiontangles}. 
\item Rational tangles imply that fermion mixings arise and are described by the usual phases and angles (see \cite{cspepan}). 

\item The tangle model implies that gauge interactions exist and are modelled  by tangle deformations. Exactly three kinds of gauge interactions follow, due to the three Reidemeister moves.
\item Tangles determine the elementary  \emph{boson} spectrum with the observed particle properties -- spin, charges, representations and masses (see \figureref{i-tm-bosons}).
\item The elementary boson spectrum deduced from tangles yields exactly three types of gauge interactions. 
\item The gauge bosons tangles imply that free gauge bosons are described by the usual free field Lagrangians (see \figureref{i-tm-bosons} and \figureref{i-gauge}).
\item Therefore, the boson propagators follow. \cite{csqed,cspepan}. 

\item The Higgs tangle implies that the Higgs is massive, has spin 0 and is described by its usual Lagrangian (see \figureref{i-tm-bosons}). %
\item The Higgs boson tangle explains the Yukawa mass terms by braid addition inside tangle families (see also \figureref{i-lagr2}).

\item As a consequence of the particle tangles, the \textit{only} Feynman vertices due to tangles are those of \figureref{i-lagr1} and \figureref{i-lagr2}. This also implies the algebraic form of the standard model Lagrangian. 
Because the lack of other elementary particles, and because of the topology of elementary bosons and fermions (including their infinite family members), \textit{no other} vertices and \textit{no other} propagators arise. 
Because of the lack of other vertices, \textit{no other} terms arise in the Lagrangian.

\item Tangle deformations imply that particle interactions are local, simply coupled, renormalizable, have the usual -- unbroken or broken -- gauge symmetries, obey the conservation of quantum numbers and show unique couplings (see \figureref{i-gauge}). %

\item The fundamental constants -- masses, mixing angles and couplings -- can be deduced and calculated from tangles (see Section \ref{massest}, references \cite{cspepan} and \cite{csorigin}). %
\end{itemize}
These results can be summarized:
\begin{quotation}
     \noindent \csrhd {The standard model, without any additions, omissions or modifications, results from rational tangles following the Planck-scale fundamental principle.}
\end{quotation}
In simple terms, the standard model results from tangles because the number of rational tangle families is limited to three generations, and because the number of gauge interactions is limited to three. %
Above all, the result can be tested:
\begin{enumerate}[resume]
   \item If any new interaction, such as technicolour or grand unification, any new symmetry, such as supersymmetry, any new elementary particle, or any new interaction vertex that \emph{differs} from the standard model is observed, the tangle model is falsified.

   \item If any deviation from the standard model Lagrangian (with massive Dirac neutrinos and PMNS mixing) is discovered at any energy scale, the strand conjecture is falsified.
\end{enumerate}
This \emph{quantitative} prediction predicts results for thousands of possible experiments -- from searches for leptoquarks, milli-charged particles, axions, and additional Higgs bosons to the observation of neutrino-less double beta decay \cite{cspepan,csorigin,csqed}. %
\blue{So far, all observations agree with the predictions.}

In summary, the central prediction of the strand conjecture in the domain of high energy physics is:
\begin{enumerate}[resume]
   \item There is \emph{no} measurable physics beyond the standard model with massive Dirac neutrinos and PMNS mixing (and beyond general relativity). 
   Measuring any \emph{new} effect beyond the standard model --  i.e., any effect apart from the fundamental constants -- at any energy scale, would invalidate the strand conjecture. %
\end{enumerate}
This prediction is possible because, in contrast to the standard model, the tangle model has nothing put in `by hand'; above all, the tangle model allows deducing the fundamental constants, as will become clear below.
Tangles thus imply that the standard model is not ugly, but \textit{beautifully simple:} it follows completely and uniquely from the fundamental principle of \figureref{i-tm-fundamental-principle}.

%-----------------------------------------------------------------------------
%-----------------------------------------------------------------------------
\section{Strands equations of motion}

There is a natural longing to describe strands with equations of motion.
So far, this seems difficult. 
First of all, strands have no physical properties.
Strands are unobservable.
So it makes no operational sense to define their position.
Secondly, strands are not objects of fixed length or of given tension.
They appear to be able to vary in length.
\blue{Furthermore, space and time are not continuous below the corrected Planck length.
As a result, there seems to be no way to deduce the future shape of strands from a previous shape.
It seems even more difficult to describe what happens when strand segments touch.
So it is not possible to describe strand evolution exactly.
Finally, strand equations of motion would be, in practice, hidden variables.}

\blue{In summary, \textit{there cannot be a strand equation of motion.}} 
On the other hand, equations of motions seem necessary to perform
simulations.  Also, a certain intuition for their motion is assumed in
several illustrations in this article.  Only a statistical
approach to strand fluctuations can overcome the difficulties and
satisfy the wish for shape calculations.  So far, it is assumed that
the precise behaviour of strands is such that wave functions arise in
a self-consistent way.  This issue is still topic of research and will
be explored in detail in a subsequent article, on the foundations of
quantum theory.

%-----------------------------------------------------------------------------
%-----------------------------------------------------------------------------
\section{What exactly is a quantum field theory?} 
\label{sec:fstqf}

The essence of quantum field theory is a topic of intense research.
The question was regularly asked by Weinberg \cite{weinberg} and by Zee \cite{zee}, and was explored by many scholars. %
The above sections deduced the following answer:
\begin{quotation}
	\noindent\csrhd In the strand conjecture, every \emph{quantum field} is a loose, fluctuating \emph{tangle} made of fluctuating strands with Planck radius.
\end{quotation}
As shown in the previous sections, the predictions about the tangle model about quantum field theory are numerous.
When the fluctuations of the strand crossings are averaged over time, they yield the \emph{wave function} for fermions or the \emph{field intensity} for bosons.
The tangle topology specifies the particle type and the quantum numbers. 
The tangle model yields continuous function of space, i.e., \emph{fields}, while allowing the \emph{counting} of discrete particles at the same time. %
The tangle model automatically yields the indistinguishability of identical particles.
The tangle model reproduces \emph{spin} and \emph{statistics}, as well as the theorem connecting the two properties.

The tangle model implies that every particle is a countable and localized \emph{excitation} of the (untangled) \emph{vacuum}: the tangling \emph{is} the excitation that forms a particle. %
Vacuum excitations automatically lead to both particles and antiparticles.
Particle tangles are \emph{rational}.
The (rational) tangle model reproduces all observed particle interactions:
\emph{particle creation} is due to tangling, particle \emph{annihilation} is due to untangling, particle \emph{absorption} is due to tangle combination, and particle \emph{emission} is due to tangle separation. %
The (rational) tangle model also implies that certain particles \emph{can transform} into each other, by braiding.
Also, particles \emph{can interact}: different tangles can be combined or a single tangle can be separated into two or three tangles. %

The tangle model reproduces the full \emph{elementary particle spectrum} as a result of tangle classification. 
The tangle model reproduces all quantum numbers and all particle mass values as consequence of topologic and geometric tangle properties. %

The tangle model implies that interactions are \emph{deformations} of tangles.
The tangle model implies that interactions are \emph{local} -- within Planck dimensions.
The tangle model implies that interactions are \emph{gauge interaction}.
The tangle model implies that Reidemeister moves \emph{restrict} the interaction spectrum to the three known gauge groups of the standard model. 
The tangle model implies that the weak interaction, in contrast to the electromagnetic and the strong interaction, \emph{violates parity}, \emph{breaks SU(2)}, and \emph{violates CP invariance}. %
The tangle model implies that fermions \emph{mix}, and that their mixing is described by unitary mixing matrices \cite{cspepan}.
The tangle model implies that couplings strengths are \emph{unique} and \emph{run} with four-momentum.

The tangle model implies that every perturbation expansion is due to more and more complex tangles arising at smaller and smaller scales.
In this way, the tangle model reproduces the perturbation expansion of conventional perturbative quantum field theory \cite{csqed}.

The tangle model implies that every quantum field evolution is described by a \emph{Lagrangian}, following the principle of least action.
(This eliminates quantum field theories without Lagrangians.) % See Yuji Tachikawa
The tangle model implies that gauge couplings are \emph{weak}, always smaller than 1.
The tangle model implies that the standard model is \emph{renormalizable}: tangles allow at most quadruple vertices.
The tangle model implies that quantum fields have an (unusual) \emph{strong coupling regime} -- in the tangle cores.
The tangle model implies that quantum fields have an \emph{approximate duality between strong and weak coupling} -- but not an exact one. %
The tangle model implies that quantum fields have \emph{topological} aspects -- but are \emph{not} topological quantum field theories. %
The tangle model of elementary particles is \emph{free of anomalies}, because the standard model is \cite{anomalies}. 
The tangle model provides the desired argument explaining why: the lack of anomalies is due to the topology of the elementary particle tangles and to the particle spectrum resulting from tangle classification. %

The tangle model implies that there is natural energy \emph{cut-off} at the Planck scale.
The tangle model implies that there is a well-defined \emph{continuum limit} -- for scales larger than the Planck length.
The tangle model also implies that all observables are \emph{local} -- within Planck dimensions.
The tangle model implies that observables obey \emph{axiomatic quantum field theory} -- within Planck dimensions and limits. 
(This terse summary should be investigated separately; the central arguments are from the above sections.)

The tangle model implies that space is \emph{continuous}, and not at all non-cummutative or fermionic. %
The tangle model implies that flat empty space-time is exactly \emph{Lorentz-invariant.}
The tangle model implies that physical space has \emph{three dimensions}, at all scales.
The tangle model predicts that there is \emph{only a single vacuum state.}

The tangle model predicts that non-perturbative effects must be compatible with the strand conjecture.
This topic needs more research. 
Some aspects are mentioned in the future article on the weak interaction.

Finding a single counter-example to any of the statements in this section directly invalidates the strand conjecture. 
In particular, strands and the tangle model imply 
\begin{enumerate}[resume]
   \item There is only one possible quantum field theory in nature: the standard model.
\end{enumerate}
Additional dimensions, additional particles, additional interactions, additional symmetries -- be they local, global, gauge, discrete, supersymmetric, or non-commutative -- and additional energy scales are all \emph{impossible} in nature. %
Above all, values for the fundamental constants that differ from the observed ones are not possible. %
In other words, the tangle model does \emph{not} allow generalizations of quantum field theory.
On the one hand, this feature of a quantum field theory is unusual.
On the other hand, this property is required and expected from any unified model.

In summary, the strand conjecture implies that \emph{in nature}, there is no further non-Abelian gauge group. 
This conclusion restricts the options in a famous puzzle of quantum field theory.
Specifically, strands imply that in nature there is no alternative to the standard model.  No modification and no extension is possible.
This is a centrals aspect of the strand conjecture.

%-----------------------------------------------------------------------------
%-----------------------------------------------------------------------------
%-----------------------------------------------------------------------------
% NP
\phantomsection
\phantomsection
\addtocontents{toc}{\vspace{1.5\baselineskip}}
\section*{Part II: Deducing and testing the predictions for the strong interaction}
\addcontentsline{toc}{section}{\textsf{\textbf{Part II: Deducing and testing the predictions for the strong interaction}}}

\noindent
In order to check the validity of the tangle model for the strong nuclear interaction, two aspects must be explored.
First, its has to be confirmed that all observations about the strong interaction are reproduced.
If this is not the case, the strand conjecture is falsified.
This includes non-perturbative effects, as mentioned later on.

Secondly, the tangle model has to be checked for all its predictions that go \textit{beyond} the standard 
model and, in particular, beyond QCD.
A number of the open issues mentioned at the beginning of this article -- about the origin of 
the particle spectrum and of the force spectrum -- have been answered.
The only topics beyond QCD that are left are the fundamental constants.
If the fundamental constants of QCD -- elementary particle masses and coupling constants -- are 
not reproduced, the strand conjecture is falsified.

%-----------------------------------------------------------------------------
%-----------------------------------------------------------------------------
%-----------------------------------------------------------------------------
%
\incepsfig{i-gluon}{0.97}{The tangle model for the gluon is illustrated. 
Top: while advancing, the encircled \emph{bulge} in the black strand is rotating constantly and thus yields a repeated, propagating slide. %
As usual in the strand conjecture, only crossing switches are observable. 
Bottom: there are 9 possible slides, i.e., 9 possible gluons.
Of these 9 possibilities, only eight are linearly independent: in the leftmost column, only two of the three slides are linearly independent.  % 
For all slides, partial slides can be defined and concatenated.
Detailed exploration (see text) shows that, with this concatenation, they form the group SU(3).
}[%
\psfrag{t1}{\small $t$}%
\psfrag{t2}{\small $t+\Delta t$}%
]
%-----------------------------------------------------------------------------
%-----------------------------------------------------------------------------
%-----------------------------------------------------------------------------
\section{\blue{The strand description of gluons and SU(3)}} 
\label{sec:basicgluons}

Strands imply that the strong interaction \cite{cspepan,csorigin} is due to \emph{third Reidemeister moves}, i.e., to slides: %
\begin{quotation}
\noindent \csrhd The strong nuclear interaction is the -- partial or complete -- \emph{transfer of a slide} when a gluon is absorbed or emitted. %
\end{quotation}
This yields a model for the propagation, the absorption and the emission of gluons.

In the strand conjecture, a gluon is a rotating and propagating slide in a (trivial) tangle of three strands, as illustrated on the top part of \figureref{i-gluon}. %
The rotating slide can also be described as a rotating \emph{bulge}.
The orientation of the propagating, rotating bulge defines the phase of the gluon.
Gluons, being topologically untangled, are \emph{massless} and are \emph{electrically neutral}. %
Because gluon cores are invariant after a rotation by $2 \pi$, gluons have \emph{spin 1}. % 
Gluons, being massless and being rotating slides, have \emph{two helicity states:} clockwise and anti-clockwise. 
All this agrees with observation. 

Gluons, or better, gluon bulges, advance through vacuum in a way that resembles a localised corkscrew on a strand advancing in a background network. %
The network is provided by the other strand segments in the gluon and in the surroundings. %
(Two aspects complicate this simple picture somewhat. 
First, a gluon usually does \emph{not} advance \emph{along} its tethers. %
Secondly, the corkscrew can also step over from one strand to a neighbouring one.) %
These details notwithstanding,
\begin{enumerate}[resume]
   \item Observing any deviation of the observed gluon propagator from that of a massless, neutral, spin-1 boson would falsify the strand conjecture. %
\end{enumerate}
(Massless gluons also imply that SU(3) is not broken, in contrast to the gauge group SU(2) of the weak interaction.)
The full set of possible slides -- rotations of the bulge by the angle $\pi$ -- are illustrated on the bottom of \figureref{i-gluon}. 
Full slides can be generalized to \emph{partial} slides with arbitrary rotation angles. %
A \emph{double} full slide is equivalent to \emph{no} slide at all. 
This implies that partial slides have periodic behaviour and can be described by angles, or \emph{phases}.
Also, partial slides can be \emph{concatenated:} the corresponding phases can be \emph{added}.
However, the behaviour of slides (the third Reidemeister moves representing gluons) is more complex than the behaviour of twists (the first Reidemeister moves representing photons), even though both are represented by rotating phases. %

The set of all twists (photons) has the topology of a circle, thus of U(1), as explored in reference \cite{csqed} and summarized in Section \ref{sec:gintpred}. 
In contrast, \figureref{i-gluon} shows that the set of all slides (gluons) has a more involved topology. 
The figure shows 9 different slides.
Each row of three slides represents an SU(2) group, as can easily be checked by concatenating any two slides from the same row: the result is the remaining slide (or its negative) in that same row. %
Each row follows the 
multiplication 
table (\ref{weaktab}) 
given above, in Section \ref{sec:gintpred}.
This behaviour defines SU(2).
The set of slides has \emph{three} such SU(2) subgroups, linearly independent of each other.

Three independent SU(2) groups are typical for SU(3) -- but are not sufficient to determine SU(3) uniquely.
Finishing the proof requires concatenating slides of different rows.
First of all, one notices that in the leftmost column, only two of the three slides are linearly independent.
In total, there are only 8 linearly independent slides.
Secondly, concatenating slides from different rows yields a combination of two other slides.
This is illustrated in \figureref{i-qcd-quad}.
\blue{Exploring these deformations in detail shows that slides 
follow the multiplication table given in Table~1. % su3mutablambda
Now, this table \emph{defines the Lie algebra SU(3)} \cite{cspepan,csorigin}. 
In the language used by physicists, the behaviour of slides under concatenation is the same as the behaviour of the Gell-Mann matrices under multiplication.}

\let\oldsqrt\sqrt
% Code originally from the web, modified to ignore extra height and depth.
% Dec 2013
\def\smallsqrt{\mathpalette\DHLhksqrt}
% This is brutal: it ignores ANY extra height and depth of the smallsqrt!
\def\DHLhksqrt#1#2{\setbox0=\hbox{$#1\oldsqrt{#2\,}$}\dimen0=\ht0
\setbox3=\hbox{\smash{\hbox{$#1\oldsqrt{#2\,}$}}}%
\advance\dimen0-0.2\ht0
\setbox2=\hbox{\vrule width 0.4pt height\ht0 depth -\dimen0}%cs: negative depth!
\setbox4=\hbox{\smash{\hbox{\vrule width 0.4pt height\ht0 depth -\dimen0}}}%
{\box3\lower0.45pt\box4}}% cs: original from the web was 0.4pt

{\footnotesize
\begin{table}[htp]
\captionsetup{width=145mm}
\small
\centering
\caption{\blue{The multiplication or concatenation table for the deformations $\lambda_{1}$ to
$\lambda_{8}$, deduced from \figureref{i-qcd-quad}, is the multiplication
table of the generators of SU(3).  
This particular table also includes the additional,
\emph{linearly dependent} elements akin to $\lambda_{3}$, namely
$\lambda_{9}=-\lambda_{3}/2 - \lambda_{8} \smallsqrt{3}/2$ and
$\lambda_{10}=-\lambda_{3}/2 + \lambda_{8} \smallsqrt{3}/2$; 
these elements are \emph{not} generators, but are used to construct
$\lambda_{8}$.\label{su3mutablambda}
The three SU(2) subgroups are generated by the triplet 
$\lambda_{1}$, $\lambda_{2}$, and $\lambda_{3}$, by the triplet
$\lambda_{4}$, $\lambda_{5}$, and $\lambda_{9}$, and by the triplet
$\lambda_{6}$, $\lambda_{7}$, and $\lambda_{10}$.
These are the three triplets illustrated in \protect\figureref{i-gluon}.
Despite first impression, one has $\lambda_{4}^2=\lambda_{5}^2=\lambda_{9}^2$
and $\lambda_{6}^2=\lambda_{7}^2=\lambda_{10}^2$.}}
\hspace*{-20mm}\vbox{\setlength\textwidth{160mm}\begin{equation*}
\setlength\textwidth{160mm}%
\scriptstyle
\def\s{\sqrt{3}}\let\l\lambda % 
\begin{array}{@{\hspace{0em}}c|c@{\hspace{0em}}c@{\hspace{0em}}c|c@{\hspace{0em}}%
c@{\hspace{0em}}c|c@{\hspace{0em}}c@{\hspace{0em}}c|c@{\hspace{0em}}}%
\scriptstyle %
%%%%% line ok
&  \l_1  &  \l_2  &  \l_3  &  \l_4   &  \l_5     &  \l_9&  \l_6  &  \l_7  &  \l_{10} &  \l_8  \\ % 
\hline 
%%%%%  - line ok
\l_1\cstabhlineup %
&   2/3  &  i\l_3 & -i\l_2 & \l_6/2 &  -i\l_6/2 &-\l_1/2& \l_4/2&-i\l_4/2& \l_1/2&\l_1/\s\\ %
\ \cstabhlinedown %
&+\l_8/\s&        &        &+i\l_7/2 &   +\l_7/2 &+i\l_2/2&+i\l_5/2 &+\l_5/2&+i\l_2/2&\\ % 
\hline 
%%%%%  - line ok
\l_2\cstabhlineup %    
& -i\l_3 &    2/3 &  i\l_1 & i\l_6/2 &   \l_6/2 &-i\l_1/2&-i\l_4/2&-\l_4/2&-i\l_1/2&\l_2/\s\\ % 
\ \cstabhlinedown %
&        &+\l_8/\s&        & -\l_7/2 &  +i\l_7/2  &-\l_2/2& +\l_5/2&-i\l_5/2 &+\l_2/2&\\ % 
\hline 
%%%%%  - line ok
\l_3\cstabhlineup %   
&  i\l_2 & -i\l_1 &    2/3 & \l_4/2 &  -i\l_4/2&-1/3-\l_3/3&-\l_6/2&i\l_6/2 &-1/3+\l_3/3&\l_3/\s\\ %
\ \cstabhlinedown %
&        &        &+\l_8/\s& +i\l_5/2 &   +\l_5/2 &+\l_9/3&-i\l_7/2 &-\l_7/2&+\l_{10}/3&  \\ % 
\hline 
%%%%%  - line ok
\l_4\cstabhlineup %
&\l_6/2&-i\l_6/2&\l_4/2   &2/3+\l_3/2& -i\l_9 & i\l_5 &\l_1/2 &i\l_1/2 &-\l_4/2 &-\l_4/2\s \\ %
\ \cstabhlinedown %
&-i\l_7/2&-\l_7/2&-i\l_5/2&-\l_8/2\s &         &      &+i\l_2/2 &-\l_2/2 &-i\l_5/2&-i\s\l_5/2\\ % 
% %
\hline 
%%%%%  - line ok
\l_5\cstabhlineup %
& i\l_6/2&\l_6/2&i\l_4/2 &   i\l_9  &2/3+\l_3/2&-i\l_4&-i\l_1/2&\l_1/2 &i\l_4/2&i\s\l_4/2 \\ %
\ \cstabhlinedown %
& +\l_7/2&-i\l_7/2&+\l_5/2 &        &-\l_8/2\s &      &+\l_2/2 &+i\l_2/2 &-\l_5/2&-\l_5/2\s\\ % 
\hline 
%%%%%  - line ok
\l_9\cstabhlineup % 
&-\l_1/2 &i\l_1/2&-1/3-\l_3/3&-i\l_5&i\l_4&2/3+2\l_3/3&\l_6/2  &i\l_6/2&-1/3-\l_9/3& -1\\ %
\ \cstabhlinedown %
&-i\l_2/2&-\l_2/2&+\l_9/3    &      &     &+\l_9/3&-i\l_7/2&+\l_7/2&+\l_{10}/3& +\l_{10}\\ %
\hline 
%%%%%   - line ok
\l_6\cstabhlineup % 
&+\l_4/2&i\l_4/2 &-\l_6/2 &   \l_1/2  &i\l_1/2&\l_6/2&2/3-\l_3/2 &i\l_{10}&-i\l_7&- \l_6/2\s\\ %
\ \cstabhlinedown %
&-i\l_5/2&+\l_5/2 &+i\l_7/2 & -i\l_2/2 &+\l_2/2&+i\l_7/2&-\l_8/2\s&        &       &-i\s\l_7/2 \\ % 
\hline 
%%%%%   - line ok
\l_7\cstabhlineup % 
&i\l_4/2 &-\l_4/2 &-i\l_6/2&  -i\l_1/2&  \l_1/2&-i\l_6/2&-i\l_{10}& 2/3-\l_3/2&i\l_6&i\s\l_6/2 \\ %
\ \cstabhlinedown %
&+\l_5/2 &+i\l_5/2 &-\l_7/2 &  -\l_2/2 &-i\l_2/2&+\l_7/2&         &-\l_8/2\s&     &-\l_7/2\s\\ %
\hline
%%%%%  - line ok
\l_{10}\cstabhlineup % 
&-\l_1/2&-i\l_1/2&-1/3+\l_3/3&-\l_4/2&-i\l_4/2&-1/3-\l_9/3&i\l_7&-i\l_6 &2/3-\l_3/3&1\\ %
\ \cstabhlinedown %
&+i\l_2/2&-\l_2/2&-\l_{10}/3&+i\l_5/2&-\l_5/2&+\l_{10}/3 &      &       &+\l_9/3   &+\l_9\\ %
\hline 
%%%%%  - line ok
\l_8\cstabhlineup %
&\l_1/\s &\l_2/\s& \l_3/\s&-\l_4/2\s&-i\s\l_4/2&-1      &-\l_6/2\s&-i\s\l_6/2& 1    &2/3 \\ %
\ \cstabhlinedown %
&        &       &        &+i\s\l_5/2&- \l_5/2\s&+\l_{10}&+i\s\l_7/2&-\l_7/2\s& +\l_9&-\l_8/\s\\ %
\end{array}%
\nonumber
\end{equation*}}\hss%  
\end{table}}

\blue{With the time averaging of crossing switches, each gluon leads to a gluon field described by four-potential $G^\mu_a$. The four-potential is the momentum density of gluon type $a$. The averaging is the same as in electromagnetism, where the vector potential is the averaged momentum density of photons \cite{csqed}. (In intuitive terms, the time averaging gets rid of the tethers and reduces the vector bosons to a crossing density due to their moving cores.) 
When the combinatorial properties of the gluons deduced from \figureref{i-qcd-quad} -- that lead to 
Table~1 -- are introduced, the free gluon Lagrangian with SU(3) gauge symmetry arises without alternative
\begin{equation}
  L_{\rm gluon}= -\frac{1}{4}G^a_{\mu\nu}G^{\mu\nu}_a \newline
  \hbox{\ with \ } G^a_{\mu\nu} = \partial^\mu G^\nu_a
  -
  \partial^\nu G^\mu_a
  -
  g\, f^{abc} \, G^\mu_b G^\nu_c
  \;\;.
  \label{gluonlag}
\end{equation}
In simple terms, \textit{slides yield SU(3).} This solves a further open problem of the standard model.}

\blue{In particular, the slides illustrated in \figureref{i-qcd-quad} determine the SU(3) structure constants $f^{abc}$ appearing in the colour field tensor $G^a_{\mu\nu}$ and fix the number of gluon tangles at 8. 
(Higher terms in the fields cannot arise, because slides only allow quadruple vertices, as explained in the next section.) 
The still undetermined value of the strong coupling constant $g^2 \sim \alpha_{\rm s}$ will be deduced below.}

% Jan 22
In summary, SU(3) is a natural result of the strand conjecture that is due to slides, the third Reidemeister moves.

%-----------------------------------------------------------------------------
%
\incepsfig{i-qcd-quad}{1}{The quadruple gluon vertex of the strong interaction in the tangle model is illustrated. 
In this process, combinations of two gluons yield combinations of two other gluons. 
The figure shows that colour is conserved in the process.}[%
\psfrag{t1}{\small $t$}%
\psfrag{t2}{\small $t+\Delta t$}%
]

%-----------------------------------------------------------------------------
%
\incepsfig{i-tm-qcd-basic}{1}{The strand conjecture for the triple verteices of quantum chromodynamics is illustrated. 
   Only crossing switches are observable.
   Top: the absorption of a gluon by a coloured strand triple in a tangle, at Planck scale. 
   Centre: the corresponding observation. 
   Bottom: the corresponding Feynman vertices.}[%
%\psfrag{t1}{\small $t$}%
%\psfrag{t2}{\small $t+\Delta t$}%
]
%-----------------------------------------------------------------------------

%-----------------------------------------------------------------------------
%-----------------------------------------------------------------------------
%-----------------------------------------------------------------------------
\section{\blue{From strands to the strong interaction and quantum chromodynamics}} % October 2021
\label{sec:fsttsi}

The description of the strong interaction as slide exchange will turn out to completely define the strong interaction and quantum chromodynamics. This section explores the interaction vertices.
As illustrated in \figureref{i-tm-QCDdiagram} at the beginning of this article,
there are three types of strong interaction processes in nature: the quark-gluon vertex, the triple gluon vertex, and the quadruple gluon vertex. 
Together, they define quantum chromodynamics.
The strand description of the three vertices is detailed in \figureref{i-tm-qcd-basic} and \figureref{i-qcd-quad}. %
The quadruple gluon vertex is a good starting point.

In nature, in the quadruple gluon vertex, a gluon \emph{pair} is absorbed and their properties are transferred to  another gluon. %
Alternatively, one can say that a gluon pair interacts and leads to two other, different gluons.
In the strand model, as illustrated in \figureref{i-qcd-quad}, a quadruple gluon vertex is a case of double slide transfer between gluons tangles.
The quadruple vertex of \figureref{i-qcd-quad} follows directly from the tangle structure of the gluons and from the impenetrability of strands.
The quadruple gluon vertex is due to the non-commutativity of SU(3), which, in the strand conjecture, is a consequence of the three strands that make up the structure of each gluon. %
Both in quantum chromodynamics and in the strand conjecture, the quadruple vertex is due to the fourth-order term in the gluon field found in the gluon Lagrangian (\ref{gluonlag}).
Interestingly, the quadruple gluon vertex has not yet been observed directly in multi-jet collision events, due to the experimental difficulties involved.
Nevertheless, quadruple gluon vertex is known to exist, because it is needed to make quantum chromodynamics consistent with measurements. %

In the two triple vertices of the strong interaction, a single gluon is \emph{absorbed} or emitted.
In the strand description, when a gluon is absorbed, the gluon slide is transferred to the tangle core of a quark or of another gluon, and a triplet of vacuum strands remains, as illustrated in \figureref{i-tm-qcd-basic}. %
At the same time, the phase of the absorbing particle changes, due to the slide that occurs there. %
In the corresponding gluon \emph{emission} process, three vacuum strands acquire a slide from the emitting tangle core. %
Also in this case, the phase of the emitting particle changes. %

\blue{The triple gluon vertex, like the quadruple gluon vertex, is already fixed by the pure gluon Lagrangian.}
It arises from the cubic term in the gluon field in Lagrangian (\ref{gluonlag})
and is again due to the non-commutativity of SU(3).
The triple gluon vertex has been observed in particle collisions already in the 1990s \cite{triplevertex}.

The most interesting vertex is the quark-gluon vertex, which describes the direct coupling of the two particles.
\figureref{i-tm-qcd-basic} illustrates that
during the slide transfer involving a quark -- i.e., during the strong interaction of a quark --  the quark \emph{phase} \emph{changes}. %
This connection reproduces the general observation that in nature, the phase of wave functions can change in exactly two ways: either by \textit{propagation} -- as described by the \emph{free} Dirac equation -- or by \textit{interaction} -- as described by the Feynman vertices or by the \emph{full} Dirac equation. %
The complexity of the slide exchange process is also the reason that the strong interaction has short range.

\blue{The slide transfer occurring in the quark-gluon vertex determines the last remaining part of the QCD Lagrangian.
Again, this occurs in the same way as in the case of QED \cite{csqed}.
The algebraic details are uniquely fixed by the tangle structures of the quarks and of the gluons.
After time averaging, the tangles lead to the usual expression of the interaction Lagrangian
\begin{equation}
L_{\rm int}=
    \sum_q^N \bar\psi_{qj}\,(i g \left( G^a_\mu \lambda_a \right)_{jk})\,\psi_{qk} 
\;\;,
\end{equation}
in which the generators $\lambda_a$ of SU(3), the Gell-Mann matrices, appear directly with the gluon fields $G^a_\mu$. The value of the strong coupling constant $g^2 \sim \alpha_{\rm s}$ will be deduced below. This completes the Lagrangian of quantum chromodynamics.}

A remark is useful at this point. In the strand description of nature, lengths of the order of the minimal length are effectively negligible.
Because a slide transfer -- i.e., the strong interaction -- arises in a volume of a few cubic Planck lengths defined by a strand triplet, the strong interaction is effectively \emph{local}. %
Because the slide transfer arises in a finite volume of extremely small, but finite size, there are \emph{no issues} with UV divergences.
Because the emission or absorption of a gluon is effectively local, the tangle model of QCD explains that a charged particle can have a spread-out wave function and nevertheless can behave as (almost) point-like in interactions. %
On the one hand, the wave function is due to the tangle fluctuations of the complete tangle, which is spread out in space.
On the other hand, the strong interaction occurs at a strand triplet, which is effectively point-like.

In summary, \blue{the tangle model of gluons and quarks describes the Feynman vertices of the strong interaction, and no additional vertices. Therefore, the tangle model reproduces the Lagrangian of quantum chromodynamics. In particular,}
\begin{enumerate}[resume]
   \item Observing any deviation from quantum chromodynamics, from the SU(3) gauge invariance, from the locality of the strong interaction (within Planck dimensions), or from the QCD Lagrangian would falsify the strand conjecture. %
\end{enumerate}
\noindent
Because only triple and quadruple vertices occur in the strong interaction, making use of the usual theorems, quantum chromodynamics is \emph{renormalizable}. %
\blue{This property thus directly follows from the strands tangle model.}

%-----------------------------------------------------------------------------
\incepsfig{i-colour}{1}{In the tangle model, the colour of a quark is due to the orientation in space, around its tail, of its three other tethers. %
The \emph{tail} of a quark is one of its four tethers and is defined in \protect\figureref{i-tm-fermiontangles}. %
In this illustration, quark tangles are drawn as consisting of a dark (blue) and a light (grey) strand. %
The tangle core is drawn as a central black dot.}
%-----------------------------------------------------------------------------

%-----------------------------------------------------------------------------
%-----------------------------------------------------------------------------
%-----------------------------------------------------------------------------
\section{Predictions about colour} % October 2021
\label{sec:6elch}

In nature, a particle is called uncoloured or \emph{white} if its phase does \emph{not} change when absorbing random gluons. %
A particle is called \emph{coloured} if its phase \emph{changes} in a preferred direction when absorbing random gluons. %

In the strand conjecture, the tangle cores of all \emph{white} elementary particles contain `\emph{even}' strands triplets, i.e., triplets that can be approached from the front and from the back. %
(In the diagrams of the leptons shown in \figureref{i-tm-fermiontangles} and \figureref{i-tm-electronpositron}, this corresponds to approaches from above and from below the paper plane.) %
As a result, white particles show \emph{no} average phase change when they are hit by random gluons. %
Examples of such cores are electron tangles, and all the other lepton tangles.
Also W, Z and Higgs bosons, illustrated in \figureref{i-tm-bosons}, are made of three strands and have even triplets.
All these particle are white.

In contrast, other elementary particles have `\emph{odd}' strands triplets that can be approached only from one (e.g., front) side. %
Such cores have a preferred rotation direction when they absorb random gluons: they are \emph{coloured}. %
The only examples are \emph{quark} tangles, shown on the top of \figureref{i-tm-fermiontangles} and in \figureref{i-3-generations}.
For quarks, the three colour options are defined by the three possible orientations in space around the quark tail of the three other tethers.
The details are shown in \figureref{i-colour}. 
(The tail of a quark is uniquely defined for all quarks except for the d quark, as illustrated in \figureref{i-tm-fermiontangles}. 
This exception explains the mass anomaly of the d quark, as detailed below.)

Still other elementary particles tangles are \emph{trivial triplets}, i.e., untangled triplets: this is the case for \emph{gluons} only. %  
They carry \emph{two} colours each, as illustrated in \figureref{i-gluon}. 
There are 8 linear independent colour combinations.

In the tangle model, colour is a property that is due to a mixture of topology and orientation in space. As a result,
\begin{enumerate}[resume]
   \item Discovering any \emph{exception} to colour charge \emph{quantization} would falsify the strand conjecture. %
\end{enumerate}
In the strand conjecture, colour is thus predicted to arise only in fermions made of two strands and in massless bosons made of three strands. %
The strand definition of \emph{quark colour} implies that it has three different possible options, that it is quantized, conserved, and, by definition, able to emit and absorb single virtual gluons. %
The amount of colour of quarks and that of antiquarks are predicted to be exactly the same, but of opposite sign. %
All this agrees with observation.

The strand definition of \emph{gluon colour} implies that it has eight options, due to combinations of three -- minus the linear dependent option -- and again that colour is quantized, conserved, and able to emit and absorb single or pairs of virtual gluons. %
As a consequence,
\begin{enumerate}[resume]
   \item Observing a \emph{coloured lepton} or a \emph{coloured elementary boson} -- other than gluons --  would falsify the strand conjecture. %
\end{enumerate}
Because colour is a \emph{topological} and \emph{geometric} consequence of strand triplets, the strand conjecture implies that in all interactions and in all Feynman diagrams \emph{colour is conserved.} %
\begin{enumerate}[resume]
   \item Discovering an exception to colour \emph{conservation} would falsify the strand conjecture.
\end{enumerate}

In summary, because colour is conserved, every coloured strand triplet can be said \emph{to carry} a conserved colour \emph{charge}. %

%-----------------------------------------------------------------------------
%-----------------------------------------------------------------------------
%-----------------------------------------------------------------------------
% October 2021, 20
\section{The coupling between matter and the colour field} 
\label{sec:coupling}

\figureref{i-tm-qcd-basic} illustrates the coupling between quarks and gluons, i.e., between matter and the colour field, in the strand tangle model.
First of all, the strands in the figure show that the coupling is proportional to colour charge: \emph{higher} colour values have more crossings and thus emit \emph{more} gluons. %
Likewise, in an absorption process, higher colour values absorb more gluons.
The coupling to the gluon field is thus proportional to colour charge  -- and in particular independent of the other particles properties such as mass. %
Therefore
\begin{enumerate}[resume]
   \item The strong coupling constant is predicted to be the same for all particles, \blue{independently of their colour values}.
\end{enumerate}
\figureref{i-tm-qcd-basic}  also confirms that the absorbed or emitted gluon changes the \emph{phase} of a colour charge \cite{cspepan}. 
More gluons have a larger effect on the phase.
Strands can  be seen as visualizing the effect of colour fields in the same way that the descriptions of Feynman \cite{feynqed}, of Hestenes \cite{hestenes,hestenes2,hestenes3} and of Baylis \cite{baylis,baylis2} did for the electromagnetic field: %
the \emph{colour field intensity} is defined by the spacetime \emph{rotation rate} that it induces on a coloured quark.
Strands realize this definition with the help of slide exchange.

Together, \figureref{i-tm-qcd-basic}  and \figureref{i-gluon} confrim that the colour interaction is due to exchange of slides, and that the phase change has SU(3) properties. %
\blue{(For example, the threefold symmetry and the three SU(2) subgroups are clearly visible. Section \ref{sec:basicgluons} provided the details.)}
Equivalently, strands and slides exchanges visualize and realize the freedom to chose the phase of a tangle; 
     this was illustrated in \figureref{i-tm-strand-crossing} in general, and in \figureref{i-gluon} and \figureref{i-tm-qcd-basic} for SU(3). %
The freedom of choosing the phase leads to an SU(3) gauge freedom. 
Strands thus imply SU(3) \textit{gauge invariance}.
In particular, strands illustrate that the coupling to the colour field is \emph{equivalent} to gauge invariance: both are due to the same geometric effects. %

All this confirms that tangles exchanging slides, i.e., third Reidemeister moves, imply the \emph{SU(3) symmetry of the strong interaction.} 
In particular, strand triplets reproduce the \emph{propagators} of gluons.
The gluon and quark tangles reproduce colour conservation
at all \emph{interaction vertices} of quantum chromodynamics. %

In summary,
\begin{enumerate}[resume]
   \item Observing \emph{any deviation} of the strong interaction from SU(3) coupling, at \emph{any} energy or scale, would falsify the strand conjecture.
\end{enumerate}
Strands thus explicitly rule out grand unification.
The last prediction does \textit{not} contradict the (corrected) Planck limits.
In electrodynamics, the finite strand diameter leads to limits for electric and magnetic fields \cite{csqed}. 
Such limits also arise for colour fields.
Again, they are given by the maximum force value $c^4/4G$ (see references \cite{sab,gib,cs2003,csmax,csprd1,csprd2} and the appendix) divided by the smallest possible colour charge.
Equivalently, strands predict that no coloured elementary particle ever experiences a force larger than $c^4/4G$.
\begin{enumerate}[resume]
   \item Observing a colour field value \emph{beyond} the corrected Planck limit, a strong force beyond $c^4/4G$, or a process with a luminosity beyond $c^5/4G$ would falsify the strand conjecture. %
\end{enumerate}
In experiments, the highest strong nuclear fields are observed inside neutron stars, in quark-gluon plasmas, and inside hadrons.
All these fields are many orders of magnitude smaller than the limit value.

%-----------------------------------------------------------------------------
\incepsfig[ht]{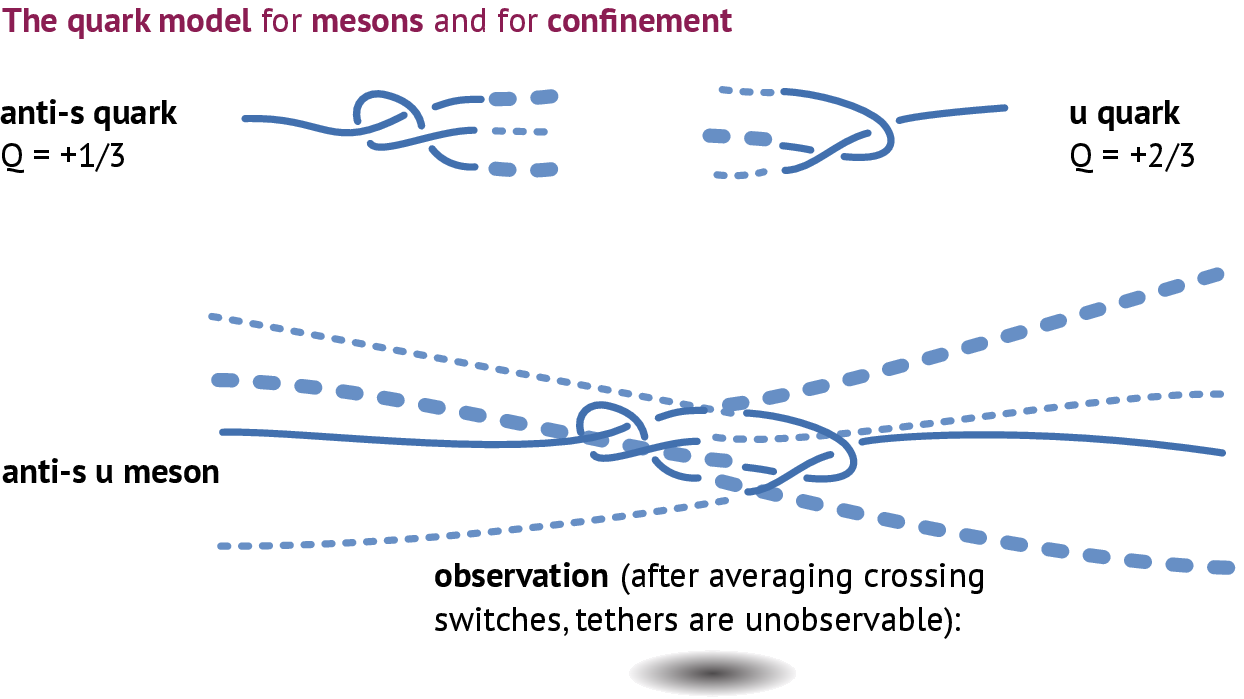}{0.9}{The quark model for a meson. 
Thick dotted tethers are \emph{above} the paper plane, thin dotted tether \emph{below}.
In a meson, two quarks of opposite colour charge attract each other.
Slides, i.e., gluons are continuously exchanged between the two quarks, i.e., between their tangle cores, yielding a \emph{colour flux tube}. %
In a meson, the tethers of the two quarks always have opposite orientation, and thus opposite colour charge: therefore all mesons are of the type $r\bar r + g\bar g + b\bar b$, and thus \emph{white}. %
At larger quark--quark distance, the fluctuations in the colour flux tube lead to more crossing switches, and thus to an energy than increases with the distance between the quarks. %
} %

%-----------------------------------------------------------------------------
\incepsfig[ht]{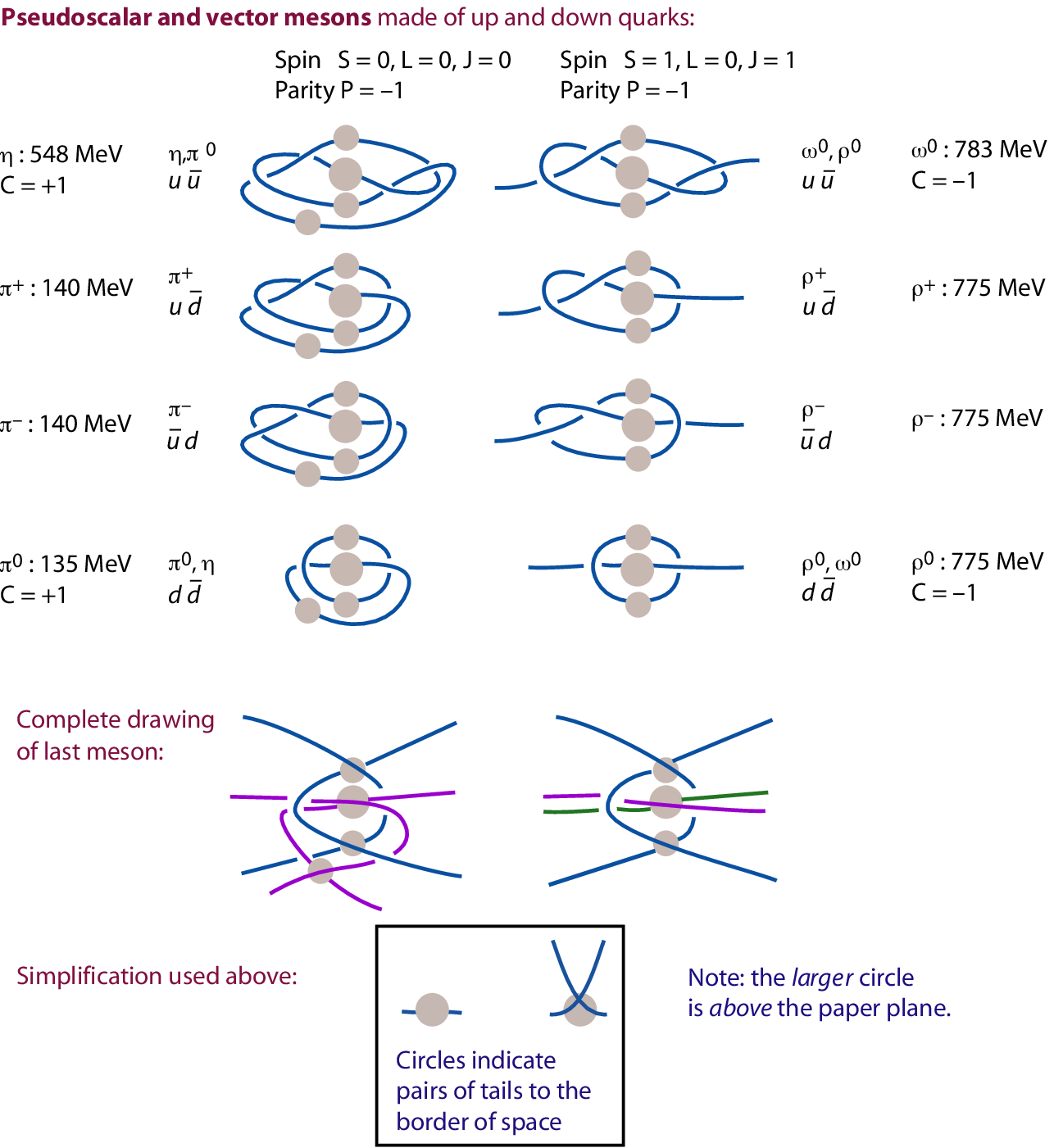}{0.9}{The simplest strand models for the light pseudoscalar 
  and vector mesons (circles indicate crossed tether pairs to the border of space), 
  with the observed mass values.}

%-----------------------------------------------------------------------------
\incepsfig[ht]{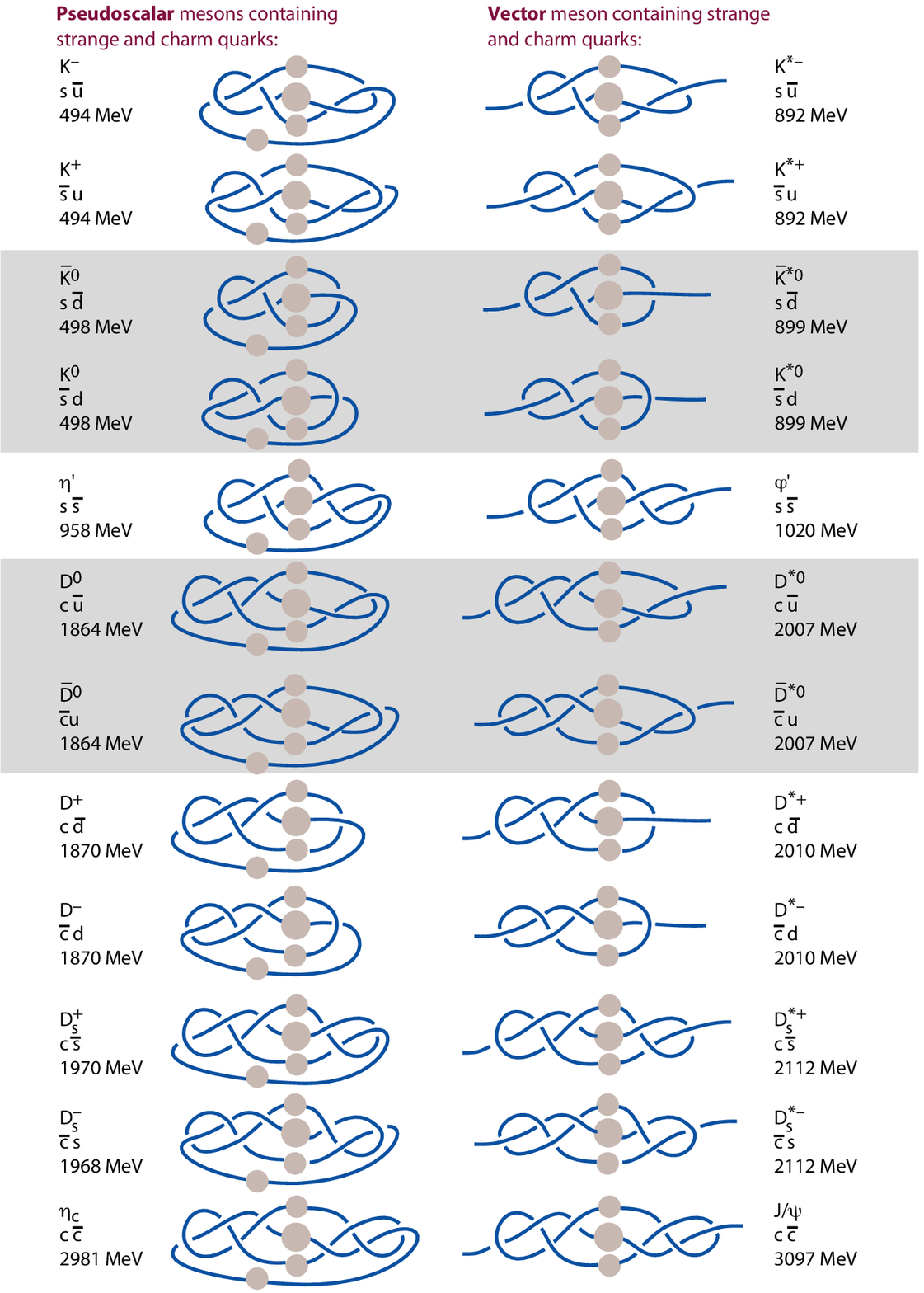}{0.9}{The simplest strand models for strange
  and charmed mesons with vanishing orbital angular momentum.  Mesons on the
  left side have spin 0 and negative parity; mesons on the right side have
  spin~1 and also negative parity. Circles indicate crossed tether pairs to the 
  border of space; grey boxes indicate tangles that mix with their antiparticles 
  and which are thus predicted to show CP violation.}

%-----------------------------------------------------------------------------
\incepsfig[ht]{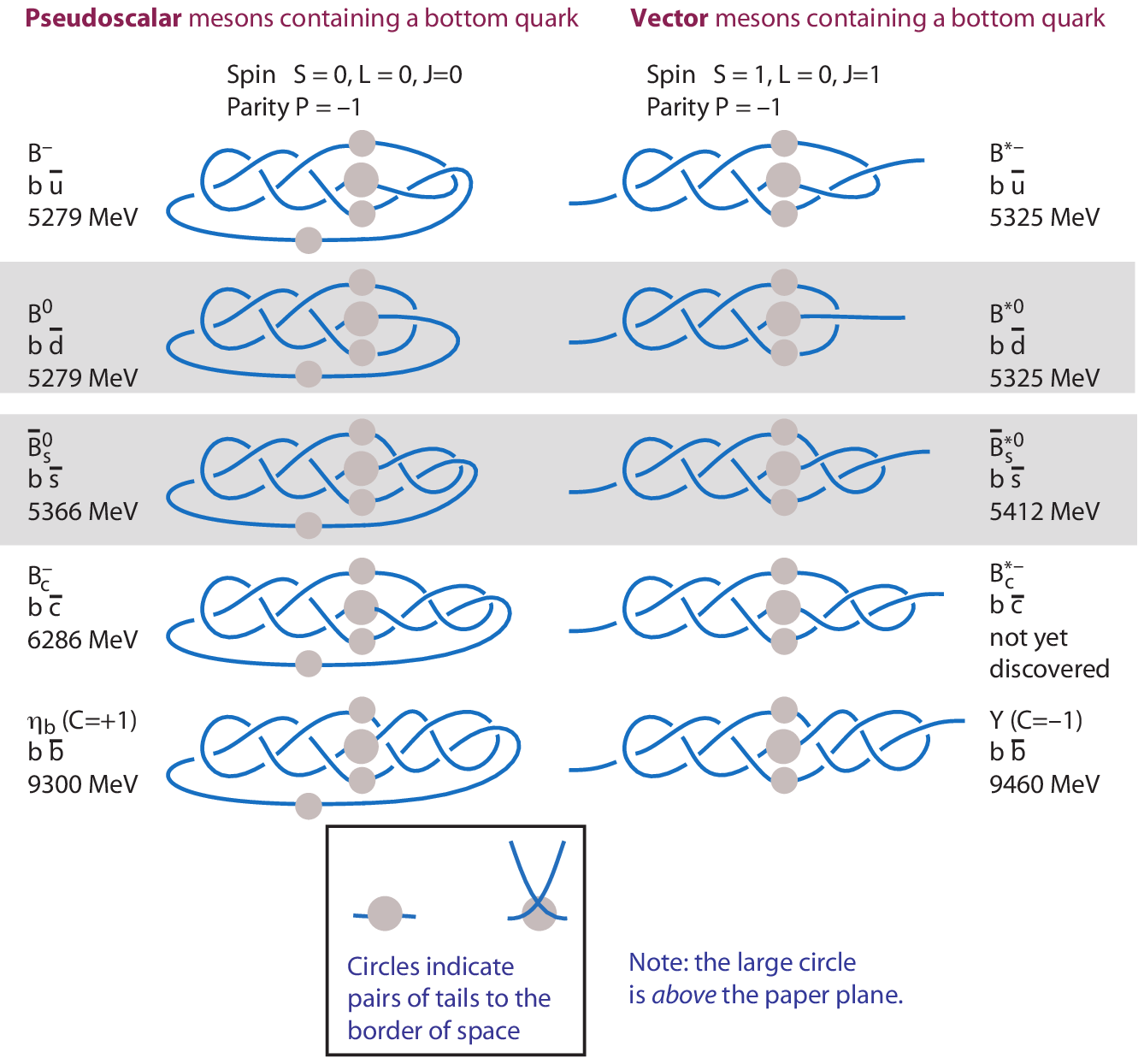}{0.9}{The simplest strand models for some heavy
  pseudoscalar and vector mesons, together with their experimental mass
  values.  Antiparticles are not drawn; their tangles are
  mirrors of the particle
  tangles.  Circles indicate crossed tether pairs to the border of space; grey
  boxes indicate tangles that mix with their antiparticles and which are thus
  predicted to show CP violation.}

%-----------------------------------------------------------------------------
%-----------------------------------------------------------------------------
%-----------------------------------------------------------------------------
% October 2021, 20
\section{From strands to the quark model for mesons and CP violation} 
\label{sec:quarkmodel}

The quark tangles in \figureref{i-tm-fermiontangles} have an important consequence that is illustrated in \figureref{i-quarkmodelmes} and in several figures that follow: 
    a quark $q$ and an anti-quark $\bar q$ can form a \emph{composite} particle.
The tangle is composite for several reasons: it can be seen as resulting from \emph{two quarks}, it is \emph{white}, and it is made of \emph{four}  strands, thus more than the maximum of three strands for elementary particles.	%
One also notes that \blue{the quark combinations} $qq$ or $\bar q \bar q$ do \textit{not} yield composite particles; such tangles do not interlock properly, as no colour flux tubes can form in those cases. %
\begin{enumerate}[resume]
   \item Observing any meson that is \emph{not white} would falsify the strand conjecture.
\end{enumerate}
So far, none were observed.

The tangle model explains all quantum numbers of mesons. 
Spin $S$ for $q\bar q$ mesons is either 0 or 1. 
For the lowest states, the orbital momentum value $L$ vanishes. (Regge states are discussed in the next section.)
Baryon number $B$ is as observed.
Electric charge follows from the chirality of the tangle and corresponds to the quark (tangle) content.
Flavour quantum numbers are obviously reproduced.
Parity P and charge parity C were introduced above, in section \ref{sec:parttruc}, and describe topological and motion symmetries of tangles.   %
In other terms, the tangle model predicts  
\begin{enumerate}[resume]
   \item Observing any $q\bar q$ meson with quantum number $J^{PC} = 0^{--}, 0^{+-}, 1^{-+}$ would falsify the strand conjecture.
\end{enumerate}
No such mesons have been observed.
This is the same argument that once led to the acceptance of the quark model of mesons.

In the tangle model, also $q\bar q q\bar q$, i.e., \emph{tetraquarks}, are possible. 
Indeed, more than a dozen such particles have been observed \cite{pdgnew}.

The tangle model in \figureref{i-quarkmodelmes} also implies that $q\bar q$ mesons are \emph{prolate}, i.e., that they all have positive quadrupole moment. %
This is observed.
In other terms, 
\begin{enumerate}[resume]
   \item Observing any oblate $q\bar q$ meson -- i.e., with a negative quadrupole moment -- would falsify the strand conjecture.
\end{enumerate}
So far, all observations confirm the prolate shape of $q\bar q$ mesons.

The meson tangles with grey background in \figureref{i-mesons2} and \figureref{i-mesons3} are the 
only ones with a topology that allows \emph{mixing} with their mirror tangles, thus with their own antiparticles. %
\blue{This is easily checked by neglecting the grey dots in the figures: Then the mirror topology is the same as the original topology, in contrast to the situation for all the other mesons.}
This leads to the prediction that
\begin{enumerate}[resume]
   \item Observing CP violation in mesons other than those marked with a grey background 
   in \figureref{i-mesons2} and \figureref{i-mesons3} will invalidate the strand conjecture. %
\end{enumerate}
So far, prediction and observations agree. 
\blue{This prediction is deduced purely from the topology of meson tangles. This again shows the explanatory power of the tangle model.}

\figureref{i-mesons1} and \figureref{i-mesons2} also illustrate the 
strand explanation of the $\eta$-$\eta^{'}$ problem.
The exact mixing  between the involved basis states is not shown. 
However, the graphs illustrate that by moving tethers around, 
one can change quark flavours and thus mix mesons.
In the strand model, the motion of tethers -- i.e., the motion of the
end-point of thethers at spatial infinity -- reproduces flavour 
mixing and meson mixing. 
This motion of tethers reproduces the non-perturbative effects usually explained with 
instantons.

The tangle model also describes the \emph{mass sequences} of mesons.  
For example, tangle complexity in \figureref{i-mesons1} predicts that the $\smash{\pi^0}$, $\eta$ 
and $\smash{\pi^{+/-}}$ have different masses and follow the observed meson mass sequence 
$\smash{m(\pi^0) < m(\pi^{+/-}) < m(\eta)}$.  
The other meson mass sequences can be checked with the help of \figureref{i-mesons1}, 
\figureref{i-mesons2} and \figureref{i-mesons3}.
Almost all mass sequences predicted from tangle complexity agree with observations.  
However, there is one unclear case: the tangle model predicts different masses for 
the $\smash{\rho^0}$, $\omega$, and $\smash{\rho^{+/-}}$ in \figureref{i-mesons1}.  
Observations confirm that the $\omega$ differs in mass from the ${\rho}$ mesons. 
Recent precision experiments can be read to suggest that $\rho^0$ and $\rho^{+/-}$ 
have different mass values, but the issue remains open \cite{pdgnew}.  %

In summary, the tangle model for quarks and mesons agrees with the traditional quark 
model concerning quantum numbers, mesons shapes, CP violation and mass sequences. %

%-----------------------------------------------------------------------------
%-----------------------------------------------------------------------------
\incepsfig{i-regge}{1}{With orbital momentum, the distance between the two quarks increases.
The higher the orbital momentum, the more crossing switches and virtual quarks can arise in 
the connecting colour flux tube region, leading to a higher mass value.} %
%-----------------------------------------------------------------------------

%-----------------------------------------------------------------------------
%-----------------------------------------------------------------------------
%-----------------------------------------------------------------------------
\section{Predictions about quark confinement and Regge behaviour in mesons} 
\label{sec:pacme}

Experiments show that like electric charge, also colour charge leads to attraction of opposite values. 
But in contrast to electric charge, colour is observed to produce a potential that increases linearly with distance (for a certain range), leading to \emph{confinement}. %
Strands explain this observation. 

For the case of \emph{mesons}, illustrated in \figureref{i-quarkmodelmes}, the tangle model implies that the distance between the quarks in a meson leads to crossing switches along the six tethers connecting the quarks. %
In other words,
\begin{enumerate}[resume]
   \item In mesons, a \emph{colour flux tube} arises. It consists of six tethers.
\end{enumerate}
Because of the larger number if strand fluctuations at larger quark distance, the effective crossing switch number per time, and thus the energy, increases with distance. %
\begin{enumerate}[resume]
   \item The tangle model of mesons thus implies an approximate potential $V \sim r$ between quarks -- as long as a flux tube is the correct description. 
   This is not the case for very short distances, when the meson tangle is too tight, nor at very large distances, when additional quark-antiquark pairs are energetically favoured.
\end{enumerate}
The linear potential between two quarks in a meson is thus a direct consequence of the tethers of the quark tangles.
In short, as a result of the tangle structure of quarks, and of the three dimensions of space, \emph{quarks are confined.}  
\begin{enumerate}[resume]
   \item The tangle model also implies that the proportionality constant in $V \sim r$ is the \emph{same} for all mesons, because the effective number of tethers \blue{in every $q \bar q$ meson flux} tube is the same. %
\end{enumerate}
The proportionality constant is indeed observed to be the same in all mesons \cite{chew,reggereview}. In addition, strands imply 
\begin{enumerate}[resume]
   \item Observing any \emph{free} and coloured particle -- quark, gluon, composite -- would falsify the strand conjecture.
\end{enumerate}

The strand model for mesons, with its flux tube, implies that the quark masses are not important for the determination of meson masses, whereas the details of the quark-antiquark bond are.  %
The fluctuations along the bond between the quarks produces additional crossing switches, and also virtual $q\bar q$ tangles, and thus increases the meson mass value. %
Experimentally, the light meson and baryon masses are indeed much higher than the masses of the constituent quarks.
For example, a $\pi$ meson, illustrated in \figureref{i-mesons1}, is much more massive than its constituent quarks.
Why is a b~quark, whose tangle is much less complex, more
massive than a $\pi$~meson?  The relation between mass and tangle
complexity only applies for tangles with the same number of tethers.
A b~quark has four tethers, a $\pi$~meson has eight.
The mass of the b~quark is mainly given by its complex tangle.
In contrast, the mass of a $\pi$~meson is mainly given by its flux 
tube.

The reduced importance of quark mass values for many meson masses is most evident for the case of mesons with a non-vanishing orbital angular momentum $L$. %
Mesons with  non-vanishing orbital angular momentum can be grouped into families which have the same quark content, but with different total angular momentum $J=L+S$.  %
These families are observed to follow \emph{Regge trajectories} when mass $m$ is plotted against total angular momentum $J$: %
\begin{equation}
    J= \alpha_{0} + \alpha_{1} m^2 \;\;,
\end{equation}
where $\alpha_{1}$ is  (almost) constant for all mesons, with a value of c.\,$0.9\,\rm GeV/fm$ \cite{chew,reggereview}. 
The \emph{Regge trajectories} derive from the linear increase of the effective potential between quarks with distance, which arises as a consequence of a sixfold fluxtube-like bond between quarks and antiquarks. %

In summary, the tangle model for mesons that arises from the  quark tangles implies  and reproduces confinement and Regge behaviour.
More research is still needed to deduce the numerical value of the common slope in Chew-Frautschi plots.

%-----------------------------------------------------------------------------
\incepsfig[ht]{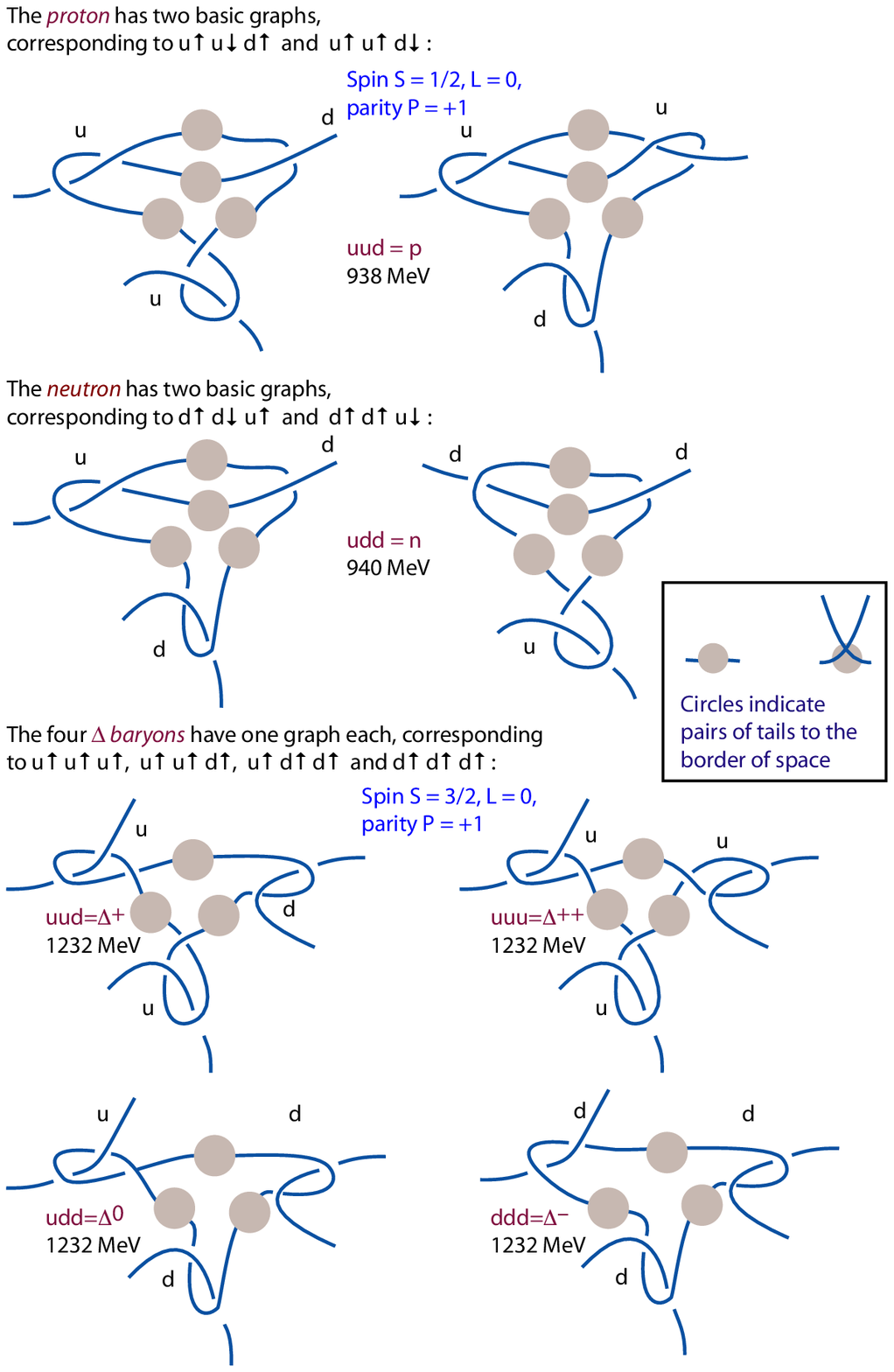}{0.9}{Baryons are made of three quarks connected by colour flux tubes.
The figure shows the simplest tangle cores of the proton and the neutron and the
  other lowest-mass
  baryons made of up and down
  quarks. Circles indicate linked tether pairs to the border of space.
  Experimental mass values are also indicated.}

%-----------------------------------------------------------------------------
%-----------------------------------------------------------------------------
%-----------------------------------------------------------------------------
\section{Predictions about baryons} 
\label{sec:paba}

Also \emph{baryons} can be explored in the same way as mesons. Many predictions about observables can be deduced,
as done in detail in reference \cite{csvol6}.
In particular, the tangle structure of quarks implies
\begin{enumerate}[resume]
   \item All $qqq$ and $\bar q\bar q\bar q $ baryons form a quark triangle; therefore, in the tangle model, all baryons are white.
 
   \item A quark or an antiquark tangle cannot `interlock' with a meson tangle, i.e., cannot form a topologically stable tangle. 
   Therefore, there are no $qq\bar q$ or $q\bar q\bar q$ baryons. 
   This is observed.
   
   \item The quark tangle model predicts exactly the same allowed and forbidden spin and $J^{PC}$ baryon quantum numbers as the usual quark model. % 
   Finding an exception would falsify the tangle model. 
   This holds also for baryon and flavour quantum numbers. 
   No forbidden quantum number baryon states arise.
   The tangle model also implies that the baryon wave function usually cannot be factorized into a spin and quark part: the nucleons
   each need two tangle graphs to describe them. %
   
   \item The tangle model predicts that in the strong interaction, quark flavour is conserved. % 
   This is observed. 

   \item The tangle model implies, like the conventional quark model, that there is a baryon singlet, an octuplet and a decuplet. 
   Tangle illustrations are found in reference \cite{csvol6}. 
   The equivalence is due to the different tangles for each quark flavour.
   Also the three-dimensional baryon classification graphs in the udsc space are recovered.
   
   \item Quark colours allow that the $\Delta^{++}$ baryon ground state wave function, which has spin 3/2, is symmetric in all three quarks. %
   Without colour, this would not be possible.
   Like in the usual quark model, also the tangle model allows the usual `way out' of the Pauli exclusion principle in baryons. %

   \item All $qqq$ baryons are oblate, i.e., have a {negative} intrinsic quadrupole moment.
   Observing any {prolate} baryon -- i.e., with a {positive} quadrupole moment -- would falsify the strand conjecture. %
   
   \item In baryons, colour flux tubes have a Y shape,  
   as illustrated in \figureref{i-y-shape}.  The prediction agrees with QCD calculations \cite{y}. 

   \item There must be \emph{common} Regge behaviour in all $qqq$ baryons, because the Y-shaped flux tubes are the \emph{same} for all baryons. % 
   Observations, for example in the $\Sigma$ and the $\Lambda$ families, confirm the prediction.
   The tangle model also implies -- because the number of strands in the flux tubes is the same -- that the slope of Chew-Frautschi plots has the same value for mesons and baryons. %

   \item Neutrons are topologically achiral, and thus electrically neutral; nevertheless neutron and anti-neutron differ.

   \item The model also naturally predicts that there are only two spin 1/2 baryons that are made of up and down quarks. 
   The two nucleons, the proton and the neutron, have similar, but different mass values.
 
   \item Tangle complexity correctly predicts baryon mass sequences among baryons with different flavour compositions.

\end{enumerate}
All this agrees with observations and expectations.
The tangle model also implies
\begin{enumerate}[resume]
   \item Pentaquarks with content $qqqq\bar q$, thus with baryon number 1, are possible.
\end{enumerate}
Observations agree with the deduction \cite{pdgnew}.
\blue{It is expected that also hexaquarks are possible.}
In other terms, the strand model states 
\begin{enumerate}[resume]
   \item Observing any deviation from the quark model with gluon composites -- such as hadrons with non-standard quantum numbers -- would falsify the strand conjecture. %
\end{enumerate}

In summary, for all baryons, the tangle model reproduces the quark model, including quark confinement, Regge behaviour, baryon mass sequences, multiquark states, \blue{and hadron quantum numbers.} %

%-----------------------------------------------------------------------------
%-----------------------------------------------------------------------------
%-----------------------------------------------------------------------------
\section{Predictions about quark mixing and CP violation} % November 2021
\label{sec:aboutqmcp}

In the strand conjecture, quark mixing and neutrino mixing occur because of the weak interaction \cite{cspepan,csorigin}.
The weak interaction exchanges tether orientations in space. 
Quark mixing is due to the braiding and unbraiding of tethers.
This deformation leads to flavour mixing.
Details will be discussed in the forthcoming article on the weak interaction. 

Strands also explain the CP violation of the weak interaction: CP violation is due to the similarity of the SU(2) group of the weak interaction and the SU(2) symmetry of spin $1/2$. %
No such possibility exists for SU(3), because slides of strands (third Reidemeister moves) do not couple to spin, in contrast to the pokes (second Reidemeister moves) of SU(2). %
Therefore, 
\begin{enumerate}[resume]
   \item There is no CP violation in the strong interaction.
\end{enumerate}
For example, strands predict a vanishing electric dipole moment of the neutron. 
So far, this is observed.

Strands also predict the lack of further elementary particles. 
Together with the automatic lack of CP violation, this implies
\begin{enumerate}[resume]
   \item There is no axion.
\end{enumerate}
In particular, the strand conjecture implies that dark matter, if it exists, does not contain axions and that axions play no role in cosmology or in astrophysics. %
So far, this prediction agrees with observations \cite{axionreview,axionreview2,axionreview3}.
If an axion is ever found, the strand conjecture is falsified.

In summary, in the tangle model, the lack of CP violation in the strong interaction is natural.

%-----------------------------------------------------------------------------
%-----------------------------------------------------------------------------
%-----------------------------------------------------------------------------
\section{Predictions about the strong coupling constant}
\label{sec:cccpred}

The value of the strong coupling constant, together with the values of the quark masses to be estimated below, determine hadron life times, hadron cross sections, branching ratios and all other experimentally accessible observables of the strong interaction. %
In quantum chromodynamics, like in quantum electrodynamics \cite{feynqed,hestenes,hestenes2,hestenes3,baylis,baylis2}, the strong coupling constant can defined in the following way: %
\begin{quotation}
\noindent \csrhd The (average) change of phase induced by the emission or absorption of a gluon 
       by a particle of unit colour charge determines the square root of the strong coupling constant. 
\end{quotation}
\noindent In the strand conjecture, the definition is the same; only the tangle model for particles is added \cite{cspepan,csorigin}. %
This reflects the situation in quantum electrodynamics: in QED, the fine structure constant is defined in the same way, and the definition agrees with that of the strand conjecture \cite{csqed}. %

Because the emission or absorption of a gluon occurs via the removal or addition of a slide, the value of the strong coupling constant is determined by the geometry of the strand process. %
Thus, the strong coupling constant can be calculated ab initio.

As explained above, strands reproduce all known Feynman diagrams, predict that no gauge groups other than U(1), SU(2) and SU(3) arise in nature, and that no new elementary particles will be discovered. %
Strands therefore predict that the standard model and its Lagrangian remain valid at all observable energy scales. %
This implies that all usual ideas and methods for the calculation of scattering and interactions remain valid, for all observable energy scales. %
\begin{enumerate}[resume]
   \item Any observed deviation from usual renormalization calculations in perturbative quantum field theory would falsify the strand conjecture. %
\end{enumerate}
In particular, the Planck-scale effects induced by the fundamental principle are so weak that they are predicted not to be observable at any experimentally accessible energy. %
Therefore, in perturbative quantum field theory, the three effective gauge coupling constants \emph{run} with four-momentum. %
In the strand conjecture, the running thus occurs just as in the standard model, where the renormalization group defines energy dependence: %
\begin{enumerate}[resume]
   \item Any observed deviation from the logarithmic \emph{running} of the coupling constants calculated with the standard model Lagrangian -- including the discovery of a new energy scale or a new symmetry --  would falsify the strand conjecture. %
\end{enumerate}
In particular, the strand conjecture reproduces the Lagrangian of QCD without modifications. 
This also implies that strands reproduce the negative beta function, and thus reproduces asymptotic freedom \cite{politzer}. 
\begin{enumerate}[resume]
   \item Discovering any difference between experiments and the beta function of QCD would falsify the strand conjecture.
\end{enumerate}
One comment can be added. 
Tangles imply the impossibility of elementary particle energy to exceed half the Planck energy, as stated in Sections \ref{sec:originsc} and \ref{sec:relqmpred}. % 
Therefore, the term `asymptotic freedom' is, strictly speaking, not correct. 
It would be more correct to speak of `Planck weakening'.
In experimental practice, however, the distinction plays no role; even the highest particle energy ever recorded in cosmic rays \cite{omg} was still over seven orders of magnitude away from the Planck energy limit. %

Strands imply that all three gauge coupling constants, including the strong coupling constant, are \emph{fixed,} \emph{unique,} \emph{calculable} and \emph{smaller than 1} \cite{cspepan,csorigin}. %
The running of the coupling constant is independent of time and space.
In particular, the strong coupling constant is predicted to be \emph{constant} over time and space -- whenever the running can be neglected. %
In addition, the strand conjecture predicts that the strong coupling constant and the other coupling constants are \emph{the same} for all particles and for all antiparticles. %
\begin{enumerate}[resume]
   \item Any experiment disproving the particle-independence, time-independence or position-independence of the running of the coupling constants would \textit{falsify} the strand conjecture. %
\end{enumerate}

But strands allows additional statements.
Slides -- the strong interaction -- require \emph{three} strands.
Twists -- the electromagnetic interaction -- require \emph{one} strand.
A slide has a stronger effect on its environment.
As a result
\begin{enumerate}[resume]
   \item The tangle model implies, because of the three spatial dimensions, that at \emph{low} four-momentum the strong coupling constant is \emph{larger} than the electromagnetic coupling constant.
\end{enumerate}
Instead, at \emph{high} four-momenta, the running of the constants reverses the situation.
In those situations, the environment is effectively lower-dimensional, and twists have more effect on their environment  than slides.
All this is observed.

In summary, strands reproduce all observed qualitative properties of the strong coupling constant.

%-----------------------------------------------------------------------------
%-----------------------------------------------------------------------------
%-----------------------------------------------------------------------------
\section{Using geometry to estimate the strong coupling constant}
\label{sec:estscc}

The slide induced by a gluon, illustrated in \figureref{i-tm-qcd-basic}, allows estimating the strong coupling constant $\alpha_{\rm s}$ ab initio.
The method used for the estimate is the same that allowed estimating the fine structure constant ab initio \cite{cspepan,csqed}. %
This is possible because in the tangle model, coupling constants are geometric in origin.
The strong coupling constant is due to the combined topological and geometric effects of slides, i.e., of third Reidemeister moves, in the same way that the fine structure constant is due to the combined topological and geometric effects of twists, i.e., of first Reidemeister moves. %

The top left of \figureref{i-tm-qcd-basic} shows a strand triangle projected on the slide triangle plane.
In the neighbourhood of the triangle, each strand is parallel to the paper plane. 
Let $\delta$ and $\epsilon$ be the angles between the sliding strand and the other two strands. %
The paper plane is best imagined as an equatorial plane.
The direction perpendicular to the paper plane is best imagined as the axis of a sphere whose north pole is above the paper and whose south pole is below it. %
The gluon incidence angle $\beta$ shown in \figureref{i-tm-qcd-basic} is a \emph{longitude} on this sphere; 
     around a given strand from a triplet it can vary from $-2\pi/6$ to $+2\pi/6$.
The other gluon incidence angle $\gamma$ is the angle from the incident gluon direction to the paper plane; 
     it thus corresponds to a \emph{latitude} and varies from $-\pi/2$ to $+\pi/2$. %

When a gluon approaches a strand triplet, it produces a complete or a partial slide. %
First, the geometric details of the photon incidence determine the \emph{value} $\nu$ of the induced phase change.% 
Secondly, the details also determine the \emph{probability} $p$ that a slide takes place. %
Finally, averaging the product $\nu p$ over all positive geometries yields the coupling constant. 
This calculation thus requires three steps.

In the first step, the \emph{value} $\nu$ of the phase change that is induced by a slide due to the incoming gluon is estimated. %
The general contribution of a strand triplet to the total tangle phase is estimated to be $(\delta +\epsilon)/2$. 
In the following, we set $\phi=\delta + \epsilon$.
A complete slide due to a gluon reverses the crossing phase from the original value to its opposite; the phase change is thus $ \nu = \phi $. %
This value is the angle by which the phase of the total tangle changes when the absorbed gluon arrives \emph{precisely} perpendicularly to the sliding strand triplet. %
The angle $\phi$ can vary from $0$ to $\pi$.
For a general gluon incidence, described by the angles $\beta$ and $\gamma$, the induced crossing switch is only \emph{partial}.
The \emph{approximate} value for the phase change is thus expected to be
\begin{equation}
\nu \approx \phi \cos \beta \cos \gamma \;\;  \;\;. %  h? % !!!no
\end{equation}
In this expression, no size dependence of the triangle has been taken into account. 
The calculation is for Planck scales only.
This approximation for phase change due to a general gluon incidence completes the first step.

In the second step, the \emph{(total) probability} $p$ that a gluon induces a slide must be estimated. 
This probability will be maximal for a gluon arriving along the poles of the incidence sphere. 
In other terms, for polar gluon incidence $\gamma=\pm\pi/2$, the probability $p$ is maximal and given by $p=1$.
In contrast, the slide probability is expected to vanish for the case $\gamma=0$, for all $\beta$ values, i.e., for \emph{equatorial} incidence. %
For general gluon incidence angles $\beta$ and $\gamma$, the switch probability $p$ also varies with the crossing angles $\delta$ and $\epsilon$; 
furthermore, the probability  varies with the gluon polarization $\zeta$ and with the gluon wavelength $\lambda$.

The \emph{slide-inducing} probability $p$ will depend on the ratio between the gluon wavelength $\lambda$ and the triangle size $h$.
The triangle size was defined in \figureref{i-tm-qcd-basic}. %
The size effect is the origin of the running of the coupling.
In the following, no size dependence of the triangle is taken into account; 
again, the calculation is for Planck scales only.

In case of polar incidence, the slide-inducing probability is estimated to be small when the triangle is obtuse,
  and large when it is acute. %  
  This suggests
	$ p \approx ( \cos \phi/2 )^2$ for polar incidence.   
The estimate has to be generalized for a general angle of gluon incidence.
The \emph{approximate} slide-inducing probability $p$ for a crossing switch is expected to change generally as %
    $ p \sim  \cos \theta_1 \cos \theta_2$, where the angles $\theta_n$ are the angles between the triangle normal and the direction of gluon incidence. %
The angles $\theta_n$ are roughly determined by the scalar products
$ \cos \theta_1 = (\cos (\phi/2), \sin (\phi/2), 0) \cdot (\cos \beta \cos \gamma , \sin \beta  \cos \gamma, \sin \gamma )$ and
$ \cos \theta_2 = (\cos (\phi/2), -\sin (\phi/2), 0) \cdot (\cos \beta \cos \gamma , \sin \beta  \cos \gamma, \sin \gamma )$. %
This yields an approximate probability 
\begin{equation}
	p \approx 
	   (\cos (\phi/2) \cos \beta \cos \gamma)^2 - (\sin (\phi/2) \sin \beta  \cos \gamma )^2
					 \;\;. 
\end{equation}
This completes the second step of the estimate of the coupling constant.

The third and final step of the calculation of the strong coupling constant is the \emph{averaging} over all possible geometries. %
The main average is over the \emph{incidence angles} $\beta$ and $\gamma$ for every absorbed gluon. %
This requires the use of the spherical surface element $(1/4 \pi ) \cos \gamma$. %
Furthermore, the calculation requires averaging over all triangle \emph{configuration angles} $\phi$ -- using the probability density for strand angles given by $\sin  \phi$. %
Each gluon can induce three different slides, yielding a factor 3. %
Finally, an average over all \emph{gluon polarizations} $\zeta$ is needed; it introduces a factor $2/3$. 
This completes the last step.

Combining the three calculation steps, the estimate for the strong coupling constant becomes %
\begin{eqnarray}
  \sqrt{\alpha_{\rm s}} \approx            % OK 
  \frac{1}{2\pi}           % CHECKED 
  \int_{\phi=0}^{\pi}      % CHECKED  
  \int_{ \beta=-2\pi/6}^{+2\pi/6}    % CHECKED    
  %\,
  \int_{\gamma=-\pi/2}^{+\pi/2}    	% CHECKED  
  \;
  p
  \;
  \nu
  \;
  \sin \phi 						%  
  \;
  \cos \gamma 						% CHECKED 
  \;
  \diffd\gamma \; \diffd\beta \; \diffd \phi   % CHECKED 
  \;\;.
  \label{alphares}
\end{eqnarray}
Inserting the above approximate expressions for $\nu$ and $p$ gives, at Planck energy, 
\begin{equation}
	\sqrt{\alpha_{\rm s}} \approx 0.13 \quad \hbox{or} \quad \alpha_{\rm s} \approx \frac{1}{61} \;\;, 
	\label{eq:tgtbt}
\end{equation}
with an \emph{unknown} error estimate.
At Planck energy, the (pure) standard model prediction is around $1/55(2)$ \cite{db1,db2}.
(At 1\,GeV, the experimental value for $\alpha_{\rm s}$ is between 1 and 1/2, depending on the renormalization scheme used \cite{pdgnew,as}.
At 100\,GeV, the experimental value for $\alpha_{\rm s}$ is $1/8(1)$  \cite{db1,db2}.)
Given the crudeness of the approximations, result (\ref{eq:tgtbt}) is not unexpected, but still disappointing.
\begin{enumerate}[resume]
   \item If a future, more precise calculation of the strong coupling constant $\alpha_{\rm s}$ based on strands disagrees with measurements, the strand conjecture is falsified.
   \item If a future, more precise calculation of the running with energy disagrees with measurements, the strand conjecture is falsified.
\end{enumerate}
This predictions correspond to those made about the fine structure constant in the article on quantum electrodynamics \cite{csqed}.

In summary, so far, the slide model for the strong interaction allows only a rough estimate of the strong coupling constant.
On the positive side, the Planck scale model for the basic QCD diagrams remains promising: it reproduces all qualitative aspects of quantum chromodynamics. % 
And the he approximate value for the strong coupling constant is ab initio, unique, constant, position-independent, and equal for all particles and antiparticles with simple colour charge. %
Similar results where already obtained for QED, with the same model and estimate.

%-----------------------------------------------------------------------------
%-----------------------------------------------------------------------------
%-----------------------------------------------------------------------------
\section[Predictions about elementary particle masses]{Predictions about elementary particle masses\protect\footnote{Improved from reference \protect\cite{csqed}.}} % March 2020, April 2020
\label{sec:parmass}

In the strand conjecture, the fundamental constants are emergent properties. 
In particular, the mass values of the elementary particles are emergent.

Mass is energy divided by $c^2$, and energy is action per time. 
In the strand conjecture, every crossing switch produces a quantum of action $\hbar$. 
The (gravitational) mass value of a fermion at rest, in units of the corrected Planck mass, is thus given by the average number of crossing switches that occur per corrected Planck time. %
\begin{quotation}

\noindent\csrhd The gravitational mass value of a fermion is due to the frequency of the spontaneous belt trick that appears 
		due to strand fluctuations \cite{cspepan,csindian}.   % cite also ae and ge
		The double tether twists generated by the belt trick correspond to virtual gravitons; the belt trick thus determines gravitational mass. %

\noindent\csrhd The inertial mass value of a fermion is due to the frequency of the spontaneous belt trick that appears 
		during core motion due to strand fluctuations.   % cite also ae and ge
        The belt trick generates a displacement and thus relates rotation and displacement.
		This relation is described by inertial mass.
		\figureref{i-tm-belttrick} gives a (pale) impression of this connection.
\end{quotation}
\noindent 
Because both mass values are due to the same mechanism, in the strand conjecture,
\begin{enumerate}[resume]
   \item Inertial and gravitational mass of particles are intrinsically \emph{equal}, for all particles, at all experimentally accessible times and places. %
   \item The mass of a particle is defined by its tangle topology and shape. %
\end{enumerate}

The connection between mass and tangle details implies a number of predictions about mass.
These predictions are specific to the strand conjecture and that can be tested even before any mass value is calculated.
\begin{enumerate}[resume]
   \item Because fermion masses are due to the belt trick frequency, particle masses are \emph{not quantized} -- in contrast to gauge charges. %
   \item Because fermion masses are due to the belt trick frequency, they are surrounded by a cloud of virtual gravitons -- which arise in their tethers -- that have spin 2, mass zero, and thus lead to a $1/r^2$ dependency of gravity in flat space \cite{csindian}. %
   (See also the appendix on gravitation.)
   \item In the strand conjecture, only \emph{localized} particle tangles have mass, i.e., only non-trivial tangles. In particular, in the strand conjecture, only fermions, W, Z and Higgs bosons interact with the Higgs field, and thus have Yukawa couplings. %
\end{enumerate}
All these predictions agree with observations.
Strands also predict that
\begin{enumerate}[resume]
   \item The mass values of all elementary particles -- due to the respective belt trick frequencies -- are \emph{positive}, \emph{fixed}, \emph{unique} and \emph{constant} in time and space, across the universe. %
\end{enumerate}
The mass value of the tangle of an elementary particle is strictly positive, because the probabilities for spontaneous tangle rotations in two opposite directions differ.
The difference is due to the lack of symmetry of tangle cores with respect to rotation: all tangle shapes are chiral.
(In fact, the simplest tangle for the down quark, shown in \figureref{i-tm-fermiontangles}, is an exception: 
     it is not chiral; its other family members are, however. %
This explains the exceptional order of u and d quark mass values, as argued below.)
\begin{enumerate}[resume]
   \item Mass values for particles and antiparticles, i.e., for tangles and mirror tangles, are predicted to be \emph{equal}.
\end{enumerate}
For a system of several particles that are non-interacting, strands imply that the total mass is the \emph{sum} of the particle masses. %
This is as expected and observed.

The probability for a belt trick is low. 
In particular, the probability for the belt trick is much lower than one crossing switch per corrected Planck time. 
Thus,
\begin{enumerate}[resume]
   \item The strand conjecture predicts that mass values $m$ for elementary particles are much smaller than the (corrected) Planck mass: 
\begin{equation}
	m \; \ll \; \sqrt{\hbar c^5/4G} ={6.1 \cdot 10^{27}}\,{\rm eV} \;\;.
\end{equation} 
\end{enumerate}
The inequality agrees with experiment, and also agrees with the \emph{maximon} concept introduced long ago by Markov \cite{maximon}.

In summary, in the strand conjecture, the low probability for the belt trick is the main reason that elementary particle masses are much smaller than the Planck mass. %
The main mass hierarchy is thus explained by the tangle model.
The hierarchy among quark masses is due to their tangle complexity, as shown below.

%-----------------------------------------------------------------------------
\incepsfig{i-glueball-loophole}{1}{\blue{Two gluons on top of each other, one rotated against the other, yield a possible loophole in the argument against glueballs, provided that the strands of the upper and lower gluon interlock, i.e., provided that the tethers of the lower (black) gluon rise above the paper plane and the tethers of the upper (grey-blue) gluon reach below the paper plane. Likewise, a stack of three gluons can also be imagined.}}

%-----------------------------------------------------------------------------
%-----------------------------------------------------------------------------
%-----------------------------------------------------------------------------
\section{\blue{Predictions about glueballs and other localized states}} % Oct 2021, Nov 2021, Jan 2022 
\label{sec:aboutMG}

\blue{Assuming that the strand conjecture is correct, one can explore the possibility of glueballs.
In a gluon, the three strands are not linked;
gluons are \textit{trivial} tangles in the sense of knot theory.
Also photons are trivial tangles, consisting of one strand only.
The strand tangle model for photons implies that there are no photon balls.
For gluons, the situation differs.}

\blue{The existence of glueballs is strongly suggested by lattice QCD, as reviewed by Lanes-Estrada \cite{latticeqcd} for low energy glueballs, or by Crede and Meyer \cite{lat1} and by Mathieu et al. \cite{lat2} for general glueballs. %
The latest experimental data also suggest the existence of glueballs \cite{klempt}, 
though the issue is not yet fully settled experimentally \cite{pdgnew}.}

\blue{In fact, \figureref{i-glueball-loophole}
shows that a strand model for a $gg$ gluon is indeed possible. 
If the gluon tethers interlock in the manner illustrated, a glueball made of two gluons arises. 
Such a gluon is expected to have the quantum numbers $J^{PC}=0^{++}$, to have white colour charge (thus to be uncoloured), non-vanishing mass, and thus to be a scalar meson. 
This agrees with expectations.
Other states with other quantum numbers also appear possible.
This agrees with expectations. 
The topic requires more study, in particular to clarify decays, cross sections, and mass values.}

\blue{The \emph{mixing} of hadrons and glueballs is also possible. 
The glueball tangles can mix with hadrons, depending on the specific tangle structures.
Also this topic still needs to be explored further.}

\blue{For topological reasons, strands do not allow composites of elementary fermions and photons.
However, the case
\emph{of hybrid mesons that include a gluon}, whether of the type $qqg$ or of another type, differs.
Numerical QCD points to the existence of such states; again, the experimental situation is unclear.
In the strand model, one can imagine that a gluon is trapped between the tethers that form the flux tube in a meson, or that a gluon interlocks with the tails of a quark. This would lead to $qg\bar q$, to $q\bar q g$ and maybe even to $qg$ states. 
Future research should clarify this topic.}

\blue{In other terms, strands imply
\begin{enumerate}[resume]
   \item \blue{There is a mass gap for SU(3) --  of finite value. The glueballs and hybrid mesons predicted by the tangle model appear to be compatible with observations.} 
\end{enumerate}
The whole topic of gluon composites requires more research. The recent discovery of the configuration of \figureref{i-glueball-loophole} is a beautiful confirmation of the explanatory power of the strand description of QCD.}

The strand conjecture also allows specific predictions about a number of further localized states discussed in the research literature. Because there is only one vacuum state, strands imply
\begin{enumerate}[resume]
   \item There are \emph{no instantons;} observing one would falsify the strand conjecture. %
\end{enumerate}
This prediction applies to both the strong and the weak interaction \cite{instantons,npqcd}.
(Though there are similar non-perturbative effects, as mentioned above.)
% Nov 2021
Furthermore, in the strand conjecture,  elementary particles are rational tangles. Therefore
\begin{enumerate}[resume]
   \item There are \emph{no} {skyrmions}, {sphalerons} or {other solitons} in quantum chromodynamics. %
\end{enumerate}
Experimental evidence for any of these states in elementary particle physics would falsify the strand conjecture \cite{skyrmion}.
In the strand conjecture, electric charge is due to tangles with topological chirality \cite{cspepan,csorigin}.
The exploration of electric charge implies that
\begin{enumerate}[resume]
   \item There are \emph{no magnetic monopoles} in the strand conjecture -- not even in quantum chromodynamics. %
\end{enumerate}
This agrees with observations so far \cite{monop}.
Strands thus put into question various approaches to QCD duality \cite{confreview,confreview2}. 
At weak coupling, descriptions with the help of vortices or with monopoles have been attempted.
Strands do not appear to provide a basis for such approaches.

In summary, the strand tangle model \blue{predicts the existence of glueballs and gluon composites,} but the lack of other unusual localized states.
Furthermore, the strand conjecture suggests why calculations in QCD are difficult: 
taking into account the shape of tangles in an analytic way is not easy. %
On the other hand, the similarity and the differences of the tangle model with other 
approaches of non-perturbative QCD suggests 
\begin{enumerate}[resume]
   \item Simulations of tangle fluctuations -- and of tether motions in 
   particular -- should allow non-perturbative 
   calculations in quantum chromodynamics. This includes the 
   situation in the early universe. 
\end{enumerate}
If such simulations disagree with experiment, the strand conjecture is falsified.
In fact, the results of this article suggest that tangles of fluctuating strands, in combination with 
modern computing power, might allow faster progress in the quest to deduce nuclear 
physics from QCD than envisaged so far \cite{qcdquest,uniconf}. %

%-----------------------------------------------------------------------------
%-----------------------------------------------------------------------------
%-----------------------------------------------------------------------------
\section{Predictions about the running of quark masses} %  

In the tangle model, the cores of particle tangles get tighter, the core axis gets more aligned 
   with the direction of motion, 
   and the core gets \emph{flatter} % (CHECK: more elongated?) % No, flatter
   at higher four-momentum or energy. %
This general effects mimics the increase of relativistic mass with energy. 
At high energy, a second effect comes into play: the diameter of 
the strands is not negligible.
The strand diameter effectively modifies the frequency of the belt 
trick and thus changes the mass value.
As a result,
\begin{enumerate}[resume]
   \item Elementary particle masses \emph{run} with four-momentum. %
\end{enumerate}
This is observed.
When a belt trick occurs at high four-momentum, fluctuations around 
the tangle core will make core rotation more probable. 
Strands thus suggest that quark mass values, in particular, should \emph{decrease} with energy. 
In the standard model, this is the case, as shown by Fusaoka et al.~and by the group of Zhou \cite{hep1,hep2,zzmass}. %

In the tangle model, all quarks have four tethers. 
This implies
\begin{enumerate}[resume]
   \item All quark masses decrease approximately \emph{by the same percentage} when four-momentum is increased. %
\end{enumerate}
This is observed \cite{hep2}: all quark masses decrease by almost the same factor over a given energy range.

Because quarks have \emph{four} tethers, whereas leptons have \emph{six}, quarks are further away from sphericity.
As a result, high energy has a more pronounced effect on quarks than on leptons.
Thus, the tangle model implies
\begin{enumerate}[resume]
   \item Quark masses decrease with four-momentum \emph{more strongly} than charged lepton masses. %
\end{enumerate}
This is observed and expected: in the standard model, lepton masses decrease by less than 10\% from everyday energy up to Planck energy, whereas up and down quark masses typically decrease by a factor 4. %
In the strand conjecture, the running of mass values will be calculable in the near future, using computer simulations that take into account tangle geometry and its fluctuations.
\begin{enumerate}[resume]
   \item Any \emph{deviation} from the mass running calculated with strands from the mass running found in experiments would falsify the strand conjecture.
\end{enumerate}
In fact, if the mass runnings calculated with the standard model and with the tangle model would differ, high-energy experiments would even allow testing the strand conjecture directly. %
However, the tangle model predicts that this will not happen.

In summary, in the strand tangle model, the running of quark masses is a geometric effect, due to the change of core shape with four-momentum. \blue{All qualitative properties of the running of masses are reproduced.}

%-----------------------------------------------------------------------------
%-----------------------------------------------------------------------------
%-----------------------------------------------------------------------------
\section[Using tangle geometry to estimate the mass of elementary particles]{Using geometry to estimate the mass of elementary particles\protect\footnote{Improved from reference \protect\cite{csqed}.}} % October 2021, Jan 2022
\label{elmassest}
\label{massest}

In the strand conjecture, the mass value of a fermion is determined by the frequency of the spontaneous belt trick that appears due to strand fluctuations.   % cite also ae and ge
Because localized tangles are \emph{chiral}, i.e., asymmetrical, the spontaneous belt trick leads to a certain average number of slides follow the multing switches per time. %
(The chirality also applies to the down quark, when the admixture of the Higgs is taken into account.)
These slides defineing switches yield an average action value per time, which defines an energy and thus a mass value \cite{csorigin}.
The frequency of the spontaneous belt trick can be estimated, and with it, particle mass. 
The frequency is due to \emph{tangle shapes}.
At present, this requires approximations.

In the strand conjecture, every massive particle is represented by a \emph{family} of tangles. 
As mentioned in Section \ref{sec:partpred} and illustrated in \figureref{i-3-generations}, the family members differ by the number of Higgs braids they contain. %
In the following, the calculations only take into account the tangle of the \emph{simplest} family member. 
The effect of the other family members -- due to higher order Higgs couplings -- is mostly neglected.
This is the first approximation.

The second approximation is the assumption that the shape of a \emph{tight} tangle is the same as the \emph{average} shape of a tangle.
Recent research by Katritch et al. suggests that this is (at least) an excellent approximation \cite{stasiak}.
In the strand conjecture, tangles can be tight or loose. 
The belt trick occurs in all cases.
For extremely loose tangles, the frequency of the belt trick is expected to be independent of the tangle size.
However, for tight tangles, the rope diameter will have an influence.

A third approximation is almost automatic.
Taking tight tangles as representative ignores the running of mass with four-momentum.
For high four-momentum values, the shape of the tangle will change, and so will its mass value.
Tight tangles determine mass values at low momentum.

In contrast to the three mentioned approximations, the use of tight tangle \emph{shapes} to determine mass values implies that the diameter of strands is \emph{not} neglected -- as it was up to this section. %
In other words, gravity is not neglected in the following. 
This is as expected: determining particle mass indeed requires taking into account both quantum effects and gravitational effects. %

Estimating the probability for the appearance of the belt trick remains \emph{a difficult geometric problem.} %
The research literature does not contain any hint towards a solution.
Researchers on polymers, on fluid vortices, on cosmic strings, on string theory, on superfluids, and on statistical knot theory have not explored the topic yet. %
The following ideas should thus be seen as tentative. % into a dark room.

Particle mass $m$ is given by the number of crossing switches per time that occur around the particle. 
For a quark made of two strands, the crossing switches are generated by the tethered rotation of the core that is illustrated in \figureref{i-tm-belttrick}.
The illustration yields the mass value \blue{formula
\begin{equation}
  m \approx p \, f \, n \;\;.
  \label{eq:mass}
\end{equation}}
In this expression, the factor $p$ describes the probability for the initial double rotation, i.e., the probability to get from the \emph{first} to the \emph{second} configuration in \figureref{i-tm-belttrick}. %
For a \emph{symmetric} core, the probability of a double rotation, whatever the orientation of the rotation axis, is expected to be equal in clockwise and anticlockwise direction. %
Therefore, for symmetric tangle cores, $p$ vanishes.
For  \emph{asymmetric} tangle cores, as is the case for all rational fermion tangles (except for the simplest d quark tangle), the factor $p$ is still expected to be quite small.
The value depends on the (averaged, three-dimensional, geometric) \emph{asymmetry} of the tangle core.
The geometric asymmetry is also the quantity that couples to the Higgs braid.
A non-vanishing asymmetry thus leads to a non-zero mass.
Also, more complex tangles will have larger values of $p$.

In expression (\ref{eq:mass}), the factor $f$ is the belt trick frequency that leads from the \emph{second} configuration in \figureref{i-tm-belttrick} to the \emph{sixth} and last configuration. % 
The factor $f$ will also be small, as the belt trick competes with the inverse double rotation of the tangle core. %
Interestingly, the small frequency $f$ is expected to be roughly \emph{scale independent}: the size of the tangle core does not play an important role. %
However, the asymmetry will help here as well; $f$ will increase with tangle complexity.

Finally, in expression (\ref{eq:mass}), the number $n$ is the average number of crossing switches per belt trick. 
The number $n$ counts the crossing switches among tethers and also the crossing switches between the tangle core and the tethers.
This number depends on the \emph{size} of the tangle core; $n$ increases with tangle core size and thus with tangle core complexity.

The proposed explanation of expression (\ref{eq:mass}) for particle  mass $m$ can be checked even before performing any calculation or estimate. 
Indeed, the explanation yields a particle mass value that is equal for particle and antiparticles, constant over space and time, and not quantized in multiples of some basic number. %
The explanation yields equal gravitational and inertial mass values.
The explanation implies mass values that run with four-momentum, i.e. with the looseness and flatness of the tangle.
The explanation implies mass values which depend, via $p$, on the Yukawa coupling to the Higgs boson.
The explanation yields, due to the factor $f$, mass values that are much smaller that the Planck mass. 
Finally, as expected, each factor $p$, $f$ and $n$ increases for more complex tangles.
Large tangles thus are more asymmetric and more massive. 

In summary, at present, a direct calculation of particle mass $m$ with equation (\ref{eq:mass}), even approximate, remains elusive. 
The main reason is the geometric complexity of the belt trick, as illustrated in \figureref{i-tm-belttrick}. 
Even the simple case of d quark rotation shown in \figureref{i-quarkrotation} is hard to handle.
However, some rough numerical statements can be made.

%-----------------------------------------------------------------------------
%-----------------------------------------------------------------------------
%-----------------------------------------------------------------------------
\incepsfig{i-tight-quarks}{0.97}{The simplest tight strand models for the quarks, as determined by Eric Rawdon and Maria Fischer.}
%-----------------------------------------------------------------------------
%-----------------------------------------------------------------------------
%-----------------------------------------------------------------------------

%-----------------------------------------------------------------------------
%-----------------------------------------------------------------------------
%-----------------------------------------------------------------------------
\section{Predictions about quark mass ratios} % October 2021, Jan 2022
\label{sec:qmass}

% 20
The strand conjecture implies that for composite particles, the mass values are determined by the tangle structure.
Generally speaking, \emph{more complex} tangles -- the number of tethers being equal -- have \emph{larger} mass. %
\begin{enumerate}[resume]
   \item In the case of the hadrons, strands predict that hadrons with higher core complexity have higher mass -- when the least massive states in the Chew-Frautschi plots, i.e., the states with vanishing angular moment are compared. %
\end{enumerate}
This prediction is verified for all mesons and baryons \cite{chew}, as explored partly above and in more detail elsewhere \cite{cspepan,csorigin}. %
In the following, we focus on the quark masses.
\begin{enumerate}[resume]
   \item In the case of the quarks, \emph{tangle complexity} predicts
   \begin{equation}
	m_{\rm d} < m_{\rm u} < m_{\rm s} < m_{\rm c} < m_{\rm b} < m_{\rm t}  \;\;.
   \end{equation} 
\end{enumerate}
\emph{This is not correct.} The first inequality is wrong. Interestingly, strands provide an explanation.

In the strand conjecture, mass values are determined by both the tangle complexity and the coupling to the Higgs.
The down quark has the lowest tangle complexity. 
On the other hand, in contrast to all other quarks, its four-fold tangle symmetry implies that its coupling to the Higgs is four times higher than for the up quark. %
(In the language of the tangle model, the down quark tangle has four possible tethers, in contrast to all other quark tangles.)
For this reason, the down mass shifts to a higher value than the up quark mass.
(Chiral symmetry breaking will be discussed further in the upcoming preprint on the weak interaction.)

The tangle model thus suggests that mass comparisons are best made among the five quarks \emph{other} than the down quark.
These five quarks have tangle simplest cores that are chiral.
The tangle cores differ among each other, when they are pulled \emph{tight}, first of all by their ropelength.

The \emph{ropelength} of a tight tangle is defined as the additional length required to tie the tangle core.
So far, knot theorists have not been able to provide an analytical expression for the ropelength value of any non-trivial tangle.
The only way to determine ropelengths is to use computer approximations. 
Results for the six simplest tight quark tangles, calculated by Eric Rawdon and Maria Fischer, are 
      illustrated in \figureref{i-tight-quarks} and numerical values are given in \tableref{quropel}.
The quark mass value will strongly change with the ropelenth $l$.
In fact, the mass value will change at least as strongly as the exponent of $l$. 
This is observed.
However, there is no simple relation between ropelength $l$ and mass.
\emph{Quark mass is not determined by ropelength alone.}

{\small
\begin{table}[t]  
%\figureversion{tabular}
\small
\centering
\caption{Quarks, electric charge, calculated ropelengths (a measure for tangle complexity only) in
units of the rope \emph{diameter} of the simplest tight quark tangles of \protect\figureref{i-tight-quarks}, 
and the experimental mass values. (The ropelength values have been determined by Eric Rawdon and Maria Fischer.){\cstabhlinemuchdown}\label{quropel}}
\setlength{\tabcolsep}{0.5mm}%\setlength{\extrarowheight}{2bp}%
\begin{tabular*}{\textwidth}{%
@{\hspace{0em}}p{28mm}@{\hspace{0em}}%
@{\extracolsep{\fill}} p{18mm}@{\hspace{0em}}%
@{\extracolsep{\fill}} p{28mm}@{\hspace{0em}}%
@{\extracolsep{\fill}} p{28mm}@{\hspace{0em}}%
@{\extracolsep{\fill}} p{28mm}@{\hspace{0em}}%
@{\extracolsep{\fill}} p{28mm}@{\hspace{0em}}}
%   
%%%%%\toprule
%
\hline
{Quark tangle} & {Electric\cstabhlineup\par charge\cstabhlinedown} & {Ropelength}  & {Difference\par to previous}  & {Observed mass} & {Observed\par mass ratio to\par previous}\\
\hline
simplest d & $-1/3$\cstabhlineup   & 1.355    &           & 4.8(8) MeV/c${}^2$  &       \\
  simplest u & $+2/3$   & 4.064    & 2.708  & 2.3(1.2) MeV/c${}^2$ &  $\approx 0.5$  \\
simplest s & $-1/3$   & 7.611    & 3.548  & 95(5) MeV/c${}^2$ & $\approx 41$ \\
  simplest c & $+2/3$   & 11.132   & 3.520  & 1.275(25) GeV/c${}^2$ & $\approx 13$ \\
simplest b & $-1/3$   & 14.686   & 3.555  & 4.18(3) GeV/c${}^2$ & $\approx 3.3$ \\
  simplest t\cstabhlinedown & $+2/3$ & 18.600 & 3.913 & 173.2(1.2) GeV/c${}^2$ & $\approx 41$ \\
\hline
\end{tabular*}
\end{table}}

The ropelength values for the simplest tight quark tangles are interesting already by themselves.
The ropelength values show, as just mentioned, that the d quark mass is not explained by ropelength at all.
In fact, also the u quark tangle has a symmetry, though a lower one than the d quark.
Also the d quark is thus expected to be more massive than expected.

The ropelength values also hint that the mass of the t quark will be much larger than that of the other quarks. 
This is observed.
\begin{enumerate}[resume]
   
   \item Comparing quarks with the same charge, the ropelength values imply that the t$/$c mass ratio should be larger than the c$/$u mass ratio. %

   \item Comparing quarks with the same difference in ropelength, one finds that the c$/$u mass ratio should be similar to the b$/$s mass ratio. %
\end{enumerate}
Both consequences agree with observations. 
But a better test is possible, inspired by \figureref{i-tight-quarks}.
\begin{enumerate}[resume]
   
   \item The tangle geometries imply that the u$/$s mass ratio should be similar to b$/$t mass ratio, because in both cases two green strands are braided in the same way.
         Indeed, the mass ratio is about 41 in both cases.
\end{enumerate}
Braiding is also related to quark mixing, and more details will be explored in the upcoming article on the weak interaction.
(Due to the exceptional status of the d quark, the braiding of the orange strand when moving from the d to the u quark cannot be compared to that moving from the c to the b quark.)

Together with ropelength, also the shape of the core determines how often the belt trick will take place per unit time.
In particular, together with the ropelength, also the \emph{aspect ratio} of the core -- i.e., the core ellipticity -- will determine the relation between the frequency of the belt trick and the shift in core position, schematically illustrated in \figureref{i-tm-belttrick}. %
That figure suggests that the mass will increase with increasing aspect ratio of the tangle core.
This is indeed observed for the upper five quarks, as \tableref{quropel} shows.
However, a more precise statement is not yet possible.
\blue{Recent research suggests that the geometric chirality of the tangle core will also play a role \cite{windchirality}.} 

In summary, tangle core complexity only reproduces quark mass ratios very crudely. 
Nevertheless, the tangle model can say more.

%-----------------------------------------------------------------------------
%-----------------------------------------------------------------------------
%-----------------------------------------------------------------------------
\section{Predictions about absolute quark masses} % October 2021, Jan 2022, March 2022

The next step in the exploration of quantum chromodynamics is to deduce estimates of \emph{absolute} quark masses.
We take expression (\ref{eq:mass}) for the mass of a tangle, $m= p \, f \, n$, and apply it to quarks,
      keeping \figureref{i-tight-quarks} in front of us. 
One notes no difference for quarks of different charge, i.e., between the left and right columns.
With the figures, estimates for the {absolute} mass values of quarks are within reach.

\begin{enumerate}[resume] 
	
% ---------------------------------
% LOWER QUARK LIMIT
% ---------------------------------

\item Strands allow deducing a \emph{lower limit} for the (bare) mass values of quarks. 
Using the equation for particle mass $m= p \, f \, n$, this implies estimating lower limits for the 
     asymmetry $p$, the belt trick frequency $f$ and the average crossing number per belt trick $n$. %

The asymmetry value $p$ is determined by the tangle asymmetry.
\emph{For the d quark,} the asymmetry in the simplest tangle vanishes; an asymmetry only arises through the Yukawa term and through the mixing with the other quarks due to the weak interaction. %
It appears hard to provide an geometric ab-initio estimate; 
the asymmetry might be of the order of the ratio between the observed quark mass and Higgs mass, thus about O(1) ppm. %

\emph{For the u, s, c, b and t quarks,} the asymmetry, and thus the probability $p$ of rotation, can be estimated ab initio from the average asymmetry of their simplest tangle cores. %
For the u quark, the asymmetry at an energy $E$ is expected to be given by about one tether switch per corrected Planck volume. 
This gives   a smallest value for the asymmetry of
\begin{equation}
  p \approx O(1) \left ( \frac{E}{E_{\rm corr. Pl}} \right )^{1/3}
  \label{eq:asy}
\end{equation}
because this describes the chirality of a tangle configuration. 
For an energy of 1\,GeV, this yields
\begin{equation}
  p  \gtrsim  10^{-6} \;\;.
  \label{eq:rotp}
\end{equation}
An error of at least two orders of magnitude is expected.
Future simulations will allow determining this value more precisely.
The asymmetry $p$ is expected to increase strongly with ropelength, because the heavier quark tangles are much `more chiral' than the lighter ones.

The belt trick frequency $f$ for a light quark can only be estimated, e.g., from \figureref{i-tm-belttrick} and \figureref{i-quarkrotation}. % 
The rarity of the process makes the mass much smaller than the Planck mass.

The frequency results from the probability that the belt trick occurs \emph{despite} competition with the backwards rotation of the core.
To occur, a tether configuration has to form \emph{four} circles, one for each tether, all with the same orientation,  around the tangle core.
The size of the four circles is less important. 
For each tether, the probability is roughly given by the probability to form a circle divided by the number 
of possible rotation axes and the ways to distribute the tethers along the two ends of each axis. %
This yields the rough approximation
\begin{equation}
  f  \gtrsim \left ( \frac{{\rm e}^{-2\pi}}{ 4 \cdot 2} \right )^4 
  \approx  3 \cdot 10^{-16} %  
  \;\;,
  \label{eq:btf}
\end{equation}
where the number of tethers determine the exponent value 4. 
The systematic error is again expected to be up to two orders of magnitude.

Given the four tethers of quarks, the number $n$ of crossing switches occurring during a quark belt trick can also be estimated.
The minimum value can be deduced by counting the crossing switches among the tethers only, neglecting those involving the tangle core. %
This gives, for four tethers, as shown in \figureref{i-tm-belttrick},
\begin{equation}
  n \geq 4 \;\;.
\end{equation}

Combining the estimates for $p$, $f$ and $u$, the lower mass bound $m$ for quarks is  
\begin{equation}
  \frac{m}{\sqrt{\hbar c^5/4G}} \geq p \, f \, n \approx 10^{-21\pm 
  4}  %  
  \;\;,
  \label{eq:ll}
\end{equation}
i.e., between 0.6\,keV$/c^2$ and 60\,GeV$/c^2$, using $\sqrt{\hbar c^5/4G}=6.1 \cdot 10^{18}\,{\rm GeV}/c^2$. % 
The range for the lower quark mass limit is compatible with the present experimental data \cite{pdgnew}, as listed in \tableref{quropel}. 
However, the difficulty of deriving a reliable lower mass limit is evident.
Nevertheless, it can be said: if a future, improved calculation of quark masses based on tangles is in contrast with data, the tangle model is \emph{falsified.}

\item Strands also allow deducing an \emph{upper limit} for the mass values of the {quarks}. %
Again, the present estimates are only crude.
The core rotation probability $p$ due to the asymmetry of the heaviest quark will be of the order $O(0.01)$, again with a large error, due to the higher inertia of heavy quarks
The estimate for the belt trick frequency $f$ will change for tangle cores that are elongated; 
     the factor $(4 \cdot 2)^4$ will then be of the order of 
	 $O(10)$.  %  
For the most massive quark, the estimate will change to $n \approx 
12$. %  

These estimates could be checked either in numerical simulations or in experiments, using tangles made of flexible silicon strings in a flowing liquid.

As a result of these guesstimates, the upper mass limit for quark mass is given by
\begin{equation}
  m_{\rm t} \approx 10^{6\pm 2}  m_{\rm d}   \;\;. %  
  \label{eq:ul}
\end{equation}
This has to be compared to the observed ratio of around $0.86 \cdot 10^{5}$. 
If a future, improved calculation of quark mass values disagrees with data, the tangle model is falsified.
There is a strong need for better approximations; though the mathematical challenge is not easy.

\end{enumerate}

The tangle model allows additional statements about elementary particle mass.
In the last century, it was common to state that a part of the mass of a charged elementary particle is so-called ``electromagnetic mass'' \cite{feyn2}. 
It was assumed that the total electromagnetic energy contained in the field around the assumed point-like charge  is a sizeable part of particle mass.
Like quantum mechanics, also the tangle model is compatible with the idea for the case for elementary particles.
In the tangle model, the electric charge -- which is due to topological tangle core chirality -- is smeared out over space. 
Since elementary particles are not point-like in the strand model, 
\blue{electromagnetic field energy is not the main part of particle mass.
Indeed, quarks suggest the lack of such a relation.
In the strand model, particle mass is best described as due to the belt trick frequency.}

As mentioned, the tangle model shows that mass is mainly influenced by tangle core complexity. 
(The relation between mass and complexity is direct only for tangles with the same numbers of tethers.)
This relation allows deducing an interesting consequence. 
\figureref{i-tm-bosons} and \figureref{i-tm-fermiontangles} show that the (charged) top quark and the (neutral) Higgs boson have tangle cores are of \emph{similar complexity and size.}
As a result,
\begin{enumerate}[resume]
   \item The tangle model implies that the top quark and the Higgs boson have similar mass values.
\end{enumerate}
This is indeed observed.

In summary, the tangle model promises to calculate all particle masses.
Finding a way to precisely calculate the probability for the belt trick of a tethered structure is the main mathematical challenge that remains.

%-----------------------------------------------------------------------------
%-----------------------------------------------------------------------------
\section{Discussion of the estimates} %  
\label{sec:discussion}

The quark mass estimates and limits deduced above are \emph{disappointing}.
The mass estimates have large error bars, essentially because the problem is a composition of several difficult mathematical challenges from 3d geometry.  %
The specific tangle topology was not taken into account yet.
The running with four-momentum and the effects of the other tangles in the electron tangle family (i.e., the effect of coupling to the Higgs) were neglected. %
As a result, the upper mass limit is not precise and exceeds the experimental top quark mass by many orders of magnitude. %
Equally, the lower mass limit is so vague that it cannot be compared to experiments yet.

Despite the disappointing quark mass limits, two  aspects remain encouraging. 
First of all, the tangle model promises to calculate mass values ab initio.
The quark mass values are unique, constant over time and space (whenever running can be neglected), positive, equal for particles and antiparticles, equal to the gravitational mass and running with four-momentum. %
The mass hierarchy between quarks and the Planck mass is explained -- without additional assumptions.
More precise estimates of quark masses appear possible with computer simulation programs. %

\begin{enumerate}[resume]
   \item The failure to reproduce, with more precise calculation methods, any one of the observed lepton, quark, W, Z or Higgs \emph{mass values}, at any single four-momentum value, would falsify the tangle model. %
\end{enumerate}
The same mixture of disappointment and encouragement occurs for the weak interaction: 
\begin{enumerate}[resume]
   \item The failure to reproduce, with more precise calculation methods, any one of the observed lepton and quark \emph{mixing angles and CP phases} -- predicted to be exclusively due to the weak interaction -- at any single four-momentum value, would falsify the tangle model. %
\end{enumerate}

The other encouraging aspect of the strand conjecture is the potential to determine, using the \emph{same} model, the coupling constants, again ab initio. %
The predicted values for the coupling constants are unique and run with four-momentum in the same way at all times and locations, for all particles and antiparticles. %
These predictions are not made by any other model.
\begin{enumerate}[resume]
   \item The failure to reproduce, with more precise calculation methods, the observed values of the (running) fine 
           structure constant $\alpha$ or of the (running) nuclear coupling constants $\alpha_{\rm s}$ and $\alpha_{\rm w}$, at any single four-momentum value, would falsify the tangle model.
\end{enumerate}

%-----------------------------------------------------------------------------
%-----------------------------------------------------------------------------
\section{Conclusion and outlook} 
\label{sec:testsc}

The strand conjecture implies the tangle model for elementary
particles:  free propagating matter particles are modelled as
advancing rotating tangle cores.  

The strand conjecture implies that the complete Lagrangian of the standard model -- 
with massive Dirac neutrinos and PMNS mixing -- follows from a mixture of topology and three-dimensional geometry. %
The strand conjecture agrees with all experiments so far, including all
those performed in the domain of the strong interaction. %
The comparison of the tangle model with the electromagnetic interaction, which is 
due to the \emph{first} Reidemeister move,  found complete agreement with experiment \cite{csqed}. %
First investigations about the weak interaction, which is due to the \emph{second}
Reidemeister move, reach the same agreement \cite{cspepan,csorigin}.

The strong nuclear interaction is
modelled as the exchange of slides, i.e., of \emph{third} Reidemeister
moves, between particles.  This description allows deducing the Lagrangian of quantum chromodynamics,
and, \emph{in addition}, allows answering various open questions about the
strong interaction: in a natural way, 
strands explain  
the number of quarks, the SU(3) gauge symmetry, CP conservation, 
the mass gap, confinement, quark masses, and the existence of a unique strong coupling constant. 
\blue{Glueballs are predicted to arise. The whole  topic of gluon composites requires more research. It allows testing the model.}

The tangle model cannot be modified without destroying the whole structure. 
The tangle model is thus \emph{easy to falsify:} if just one conclusion, 
\blue{just one of the over ninety listed tests,} or just one prediction correctly deduced from the strand conjecture is wrong, the model must be abandoned. %

Because experiments in particle physics are not able to approach Planck scales, 
the proposed tangle structure of particles cannot be tested directly. %
However, indirect tests are possible. 
Strands deduce the \emph{lack of measurable deviations of any kind} from the standard model, \blue{with massive Dirac neutrinos with PMNS mixing.} 
Strands predict the lack of any new particles, any new symmetries, any new interactions, and any trans-Planckian effect. %
Strands predict that the \emph{only} observable effects beyond the standard model 
are the particle masses, couplings and mixing angles. %
Therefore, the strictest and the \emph{only decisive} tests for the strand 
conjecture are \emph{ab initio} calculations of the particle masses, the coupling constants and the mixing angles. %
Because present estimates are not in contrast to measurements, but not yet precise enough, the calculations still need to be improved. %
At present, the strand conjecture is one of the few proposals in the research literature allowing such calculations. 

%-----------------------------------------------------------------------------
%-----------------------------------------------------------------------------
\section{Acknowledgments and declarations}  
\label{sec:ack}

The author thanks Eric Rawdon and Maria Fischer for their ropelength calculations, 
      Jason Hise and Antonio Martos for their animations, as well as David Hestenes, Martin Haft, 
	  David Broadhurst, Volodimir Simulik, Claus Ernst, Thomas Racey, Isabella Borgogelli Avveduti, 
	  Peter Schiller, Masafumi Ata, Yuanan Diao, Jason Cantarella, 
	  Ralf Metzler, Andrzej Stasiak \blue{and an anonymous referee} for discussions. %
There are no additional data available for this work.
Part of this work was supported by a grant of the Klaus Tschira Foundation.
The author declares that he has no conflict of interest and no competing interests.

%-----------------------------------------------------------------------------
\incepsfig{i-graviton}{0.9}{\blue{Left: the strand model for a black hole illustrates the origin of black hole 
entropy. 
Right: the 
strand model for the graviton illustrates that its spin value is 2 and that it is a boson. Virtual gravitons arise during the belt trick illustrated in \protect\figureref{i-tm-belttrick}.}}
%-----------------------------------------------------------------------------

%-----------------------------------------------------------------------------
%-----------------------------------------------------------------------------

\section*{\blue{Appendix: General relativity deduced from strands}}  
\label{sec:grav}
\addcontentsline{toc}{section}{\blue{\textsf{\textbf{Appendix: General relativity deduced from strands}}}}

\blue{Einstein's field equations, and thus all the observed properties of gravitation, can be deduced from the strand conjecture in at least two ways.}

\blue{The first way is to recall that the fundamental principle of the strand tangle model states that strands have Planck radius, that the fastest crossing switch takes a (corrected) Planck time, and that each crossing switch generates an action value $\hbar$.
As a consequence, strands cannot generate a force value larger than $F_{\rm max}= W_{\rm min}/(l_{\rm min}t_{\rm min}) = c^4/4G$. 
A maximum force value implies the existence of spatial \textit{curvature}.
In fact, a maximum force value can only arise at black hole horizons. 
Now, a maximum force value $c^4/4G$ leads to Einstein's field equations, as explained in various publications in the past decades \cite{cs2003,csmax,gib,csprd1,csprd2}. 
Maximum force also implies the hoop conjecture \cite{cshoop}.}

\blue{The second way to deduce Einstein's field equations is to explore the model of black holes provided by strands:
black hole horizons are weaves of strands in which all tethers leave away from the black hole, as illustrated in \figureref{i-graviton}.
Calculating the entropy of this configuration -- due to the many crossing switches possible -- yields the Bekenstein-Hawking entropy $S= (k c^3/4G \hbar) A $, as well as the temperature of black holes \cite{csindian,csbh}.
Using the entropy and the temperature of black holes, one can then repeat Jacobson's derivation of the field equations \cite{jacobson}.
In this beautiful derivation, the entropy and temperature of black holes are used to determine the first law of horizon mechanics.
The first law in turn implies the full field equations, including the cosmological constant term. 
The strand tangle model thus yields general relativity, for all distances larger than the Planck length, without any deviation.
Only the value of the cosmological constant remains undetermined.}

\blue{The strand tangle model also implies a natural model for the graviton, shown on the right of \figureref{i-graviton}. 
With the model of the graviton, the belt trick illustrated in \figureref{i-tm-belttrick} can be seen as producing virtual gravitons on the tethers of a particle tangle.
The belt trick thus illustrates that massive particles have a gravitational field around them. 
The belt trick also implies, for the case that curvature can be neglected, that the field changes as the inverse square of the distance. Details are found in reference \cite{csindian}.}

\blue{In summary, strands reproduce curvature, maximum force, the field equations, the hoop conjecture, black hole entropy, gravitational mass, the graviton, cosmology, as well as inverse square gravity whenever spatial curvature is negligible.}

%-----------------------------------------------------------------------------
%-----------------------------------------------------------------------------

%-----------------------------------------------------------------------------

\end{document}